\documentclass[pre,aps,preprint,showpacs]{revtex4}
\usepackage{amssymb}
\usepackage{latexsym}
\usepackage{amsfonts}
\usepackage{graphicx}
\usepackage{epsfig}
\usepackage{amsmath}
\usepackage{float}     
\usepackage{verbatim}

\begin{document}

Contribution to the special issue: Statistical Mechanics and Social Sciences edited by Sidney Redner.

\title{Non-consensus opinion models on complex networks}

\author{Qian Li$^1$}\email{liqian@bu.edu}

\author{Lidia A. Braunstein$^{2,1}$}

\author{Huijuan Wang$^{3,1}$}

\author{Jia Shao$^1$}

\author{H.~Eugene Stanley$^1$}

\author{Shlomo Havlin$^4$}

\affiliation{$^1$Department of Physics and Center for Polymer Studies,
 Boston University, Boston, MA 02215, USA \\ $^2$Instituto de
 Investigaciones F\'isicas de Mar del Plata (IFIMAR)-Departamento de
 F\'isica, Facultad de Ciencias Exactas y Naturales, Universidad
 Nacional de Mar del Plata-CONICET, Funes 3350, (7600) Mar del Plata,
 Argentina.\\$^3$Delft University of Technology, Faculty of
 Electrical Engineering, Mathematics and Computer Science, 2628 CD,
 Delft, The Netherlands\\ $^4$Department of Physics, Bar Ilan
 University, Ramat Gan, Israel}

\date{10 September 2012 --- lbwssh10sep.tex}

\begin{abstract}

Social dynamic opinion models have been widely studied to understand
how interactions among individuals cause opinions to evolve. Most
opinion models that utilize spin interaction models usually produce a
consensus steady state in which only one opinion exists. Because in
reality different opinions usually coexist, we focus on non-consensus
opinion models in which above a certain threshold two opinions coexist
in a stable relationship. We revisit and extend the non-consensus
opinion (NCO) model introduced by Shao {\it et al.}\cite{NCO}. The NCO
model in random networks displays a second order phase transition that
belongs to regular mean field percolation and is characterized by the
appearance (above a certain threshold) of a large spanning cluster of
the minority opinion. We generalize the NCO model by adding a weight
factor $W$ to individual's own opinion when determining its future
opinion (NCO$W$ model). We find that as $W$ increases the minority
opinion holders tend to form stable clusters with a smaller initial
minority fraction compared to the NCO model. We also revisit another
non-consensus opinion model based on the NCO model, the inflexible
contrarian opinion (ICO) model \cite{ICO}, which introduces inflexible
contrarians to model a competition between two opinions in the steady
state. Inflexible contrarians are individuals that never change their
own opinion but may influence opinions of others. To place the
inflexible contarians in the ICO model we use two different
strategies, random placement and one in which high-degree nodes are
targeted. In both strategies, the inflexible contrarians effectively
decrease the size of the largest cluster of the rival opinion but the
effect is more pronounced under the targeted method. All of the above
models have previously been explored in terms of a single
network. However human communities are rarely isolated, instead are
usually interconnected. Because opinions propagate not only within
single networks but also between networks, and because the rules of
opinion formation within a network may differ from those between
networks, we study here the opinion dynamics in coupled networks. Each
network represents a social group or community and the interdependent
links joining individuals from different networks may be social ties
that are unusually strong, e.g., married couples. We apply the
non-consensus opinion (NCO) rule on each individual network and the
global majority rule on interdependent pairs such that two
interdependent agents with different opinions will, due to the
influence of mass media, follow the majority opinion of the entire
population. The opinion interactions within each network and the
interdependent links across networks interlace periodically until a
steady state is reached. We find that the interdependent links
effectively force the system from a second order phase transition,
which is characteristic of the NCO model on a single network, to a
hybrid phase transition, i.e., a mix of second-order and abrupt
jump-like transitions that ultimately becomes, as we increase the
percentage of interdependent agents, a pure abrupt transition. We
conclude that for the NCO model on coupled networks, interactions
through interdependent links could push the non-consensus opinion type
model to a consensus opinion type model, which mimics the reality that
increased mass communication causes people to hold opinions that are
increasingly similar. We also find that the effect of interdependent
links is more pronounced in interdependent scale free networks than in
interdependent Erd\"{o}s R\'{e}nyi networks.

\end{abstract}

\pacs{}

\maketitle

\section{Introduction}

Statistical physics methods have been successfully applied to
understand the cooperative behavior of complex interactions between
microscopic entities at a macroscopic level. In recent decades many
research fields, such as biology, ecology, economics, and sociology,
have used concepts and tools from statistical mechanics to better
understand the collective behavior of different systems either in
individual scientific fields or in some combination of
interdisciplinary fields. Recently the application of statistical
physics to social phenomena, and opinion dynamics in particular, has
attracted the attention of an increasing number of
physicists. Statistical physics can be used to explore an important
question in opinion dynamics: how can interactions between individuals
create order in a situation that is initially disordered? Order in
this social science context means agreement, and disorder means
disagreement. The transition from a disordered state to a macroscopic
ordered state is a familiar territory in traditional statistical
physics, and tools such as Ising spin models are often used to explore
this kind of transition. Another significant aspect present in social
dynamics is the topology of the substrate in which a process
evolves. This topology describes the relationships between individuals
by identifying, e.g., friendship pairs and interaction
frequencies. Researchers have mapped the topology of social
connections onto complex networks in which the nodes represent agents
and the links represent the interactions between agents
\cite{socialCastellano,socialGalam, Boc_01, Dor_02, Pas_01, ER1,
  ER1959, Bollobas, SF1, Cohen, SF, Complexnetwork,
  Newmanbook}. Various versions of opinion models based on spin models
have been proposed and studied, such as the Sznajd model
\cite{Sznajd}, the voter model \cite{voter1,voter2}, the majority rule
model \cite{majority1,majority2}, and the social impact model
\cite{social1, social2}.

Almost all spin-like opinion models mentioned above are based on
short-range interactions that reach an ordered steady state, with a
consensus opinion that can be described as a consensus opinion
model. However, in real life different opinions are mostly present and
coexist. In a presidential election in a country with two political
parties in which each party has its own candidate, for example, a
majority opinion and a minority opinion coexist. The opinions among
the voters differ, with one fraction of the voters supporting one
candidate and the rest supporting the other, and rarely will the two
opinions reach consensus. This reality has motivated scientists to
explore opinion models that are more realistic, ones in which two
opinions can stably coexist. Shao {\it et al.} \cite{NCO} proposed a
nonconsensus opinion (NCO) model that achieves a steady state with two
opinions coexisting. Unlike the majority rule model and the voter
model in which the dynamic of an agent's opinion is not influenced by
the agent's own current opinion but only by its neighbors, the NCO
model assumes that during the opinion formation process an agent's
opinion is influenced by {\it both\/} its own current opinion and the
opinions of friends, modeled as nearest-neighbors in a network.  This
NCO model begins with a disordered state with a fraction $f$ of
$\sigma_+$ opinion and a fraction $1-f$ of $\sigma_-$ opinion
distributed randomly on the nodes of a network. Through interactions
the two opinions compete and reach a non-consensus stable state with
clusters of $\sigma_+$ and $\sigma_-$ opinions. In the NCO model, at
each time step each node adopts the majority opinion of its
``neighborhood'', which consists of the node's nearest neighbors and
itself. When there is a tie, the node does not change its opinion. The
NCO model takes each node's own current opinion into consideration,
and this is a critical condition for reaching a nonconsensus steady
state. Beginning with a random initial condition, this novel
nontrivial stable state in which both majority and minority opinions
coexist is achieved after a relatively short sequence of time steps in
the dynamic process. The NCO model has a smooth phase transition with
the control parameter $f$. Below a critical threshold $f_c$, only the
majority opinion exists. Above $f_c$, minorities can form large
spanning clusters across the total population of size $N$. Using
simulations, Shao {\it et al.}  \cite{NCO} suggested that the smooth
phase transition in the NCO model in random networks is of the same
universality class as regular mean field (MF) percolation. But
simulations of the NCO model in Euclidean lattices suggest that the
process might belong to the universality class of invasion percolation
with trapping (TIP) \cite{NCO, NCOPRESS}. Apparently this is the first
time, to the best of our knowledge, that a social dynamic model has
been mapped to percolation, an important tool in statistical
physics. However, the nature of this percolation on $2D$ lattice is
still under debate \cite{NCOPRESS, Grassberger}. Exact solutions of
the NCO model in one dimension and in a Cayley tree have been
developed by Ben-Avraham \cite{NCO_solution}.

Here we present simulations suggesting that the behavior of the NCO
model, in which two opinions coexist, disappears when the average
network degree increases. When the average degree of a network is
high, the agent's own opinion becomes less effective and the NCO model
converges to the majority voter model. This was argued analytically by
Roca \cite{Percolate_or_die} and claimed also by Sattari {\it et al}
\cite{Grassberger}. In the present paper, we also generalize the NCO
model and create a nonconsensus opinion model by adding a weight
(NCO$W$ model) to an agent's own opinion. The weight $W \ge 1$
represents the strength of an individual's own opinion. Note that in
the NCO model $W$ is assumed to be $1$ like the weight of its
neighbors opinion. We find that the NCO$W$ model inherits all the
features of the NCO model, except that the critical threshold $f_c$ of
the NCO$W$ model with $W>1$ decreases when $W$ increases. This means
that strengthening one's own opinion helps smaller minority opinion
groups to survive.

The NCO model reaches a steady state in which the two opinions
coexist. This is only partially realistic. In real life, two opinions
do not simply coexist---they continue to compete. Real-world examples
include the decades-long competition between the Windows and Macintosh
operating systems and between Republicans and Democrats in US
presidential politics. All the participants in these competitions have
the same goal: winning. In order to increase their prospects of
winning, they need as many supporters (or customers) as
possible. Thus, it is interesting to study how two opinions continue
to compete after they have reached a steady state. In order to
consider both aspects, the nonconsensus steady state and the
competition, Li {\it et al.}  \cite{ICO} proposed an inflexible
contrarian opinion (ICO) model in which a fraction $\phi$ of
inflexible contrarians are introduced into the final steady state of
the NCO model and two different competition strategies are then
applied. The concepts of inflexible agents and contrarian agents were
introduced by Galam \cite{contrarian, galam1} in his work on opinion
models. In the ICO model, an inflexible contrarian is an agent that
holds an opinion contrary to that held by the majority of its
surrounding group and its opinion is not influenced by its surrounding
group---it never changes. Inflexible contrarians have one goal: to
change the opinion of the current supporters in the rival group. We
see this strategy when, for example, companies send a free product to
potential customers in order to convince them to adopt the product and
influence their friends to do the same. We study the ICO model in
order to determine, for example, whether these free products actually
do help to win the competition, how many free products are needed to
be sent, and who are the best candidates to receive the free
product. Reference~\cite{ICO} presents two strategies for introducing
inflexible contrarians into the steady state of the $\sigma_+$ opinion
groups: (i) the random strategy and (ii) the targeted (high degree)
strategy. Using these strategies, we find that the relative size of
the largest cluster in state $\sigma_+$ undergoes a second-order phase
transition at a critical fraction of inflexible contrarians $\phi_c$
below which the two opinions can coexist and above which only
$\sigma_-$ exists. Thus the ICO also belongs to the type of
nonconsensus opinion models. The results also indicate that the
largest cluster in state $\sigma_+$ undergoes a second order phase
transition that can be mapped into MF percolation similar to the NCO
model.

All opinion models described above have been studied on a single
network. However, in real social opinion dynamics, individuals
belonging to different social communities can as well communicate. In
a traditional agrarian village, for example, two separate working
relationship networks often form. Men work in the fields with other
men and women work in their homes with other women. Marriages between
men and women in this setting create interdependencies between the two
working relationship networks. As far as we know, there has been no
model study of how this kind of strong social connection between two
such different groups influences the exchange of opinions. In studying
the opinion dynamics across different groups we utilize a concept that
has recently gained wide attention: the resilience of interdependent
networks to cascading failures \cite{couple, parshani, gaonature,
  gaoprl, weiprl, Bashan, Leicht, Brummitt}. Connecting two networks
together with interdependent links allows individuals to exchange
opinions between networks. In our model, two nodes from different
networks that are connected by interdependent links represent a pair
of nodes that have strong social relations. In interdependent networks
we usually distinguish between the connectivity links between agents
within each network or community and interdependent links between
agents from different networks.

To study the effect of interdependent links on opinion dynamics, we
propose a non-consensus opinion (NCO) model on coupled networks in
which we assume different opinion formation rules for internal
connectivity and interdependent links. We assume that during the
dynamic process of opinion formation the agents that are connected
with interdependent links will have the same opinion, this being the
case because their social relationship is strong. In our model, the
NCO rules are applied in each individual network. For the coupled
pairs the following rule is applied: if two interdependent nodes have
the same opinion, they will keep this opinion, but if they have
different opinions, they will follow the majority opinion of the
interdependent network system (global majority rule). Many other
possible rules could be tested for the interdependent pairs, but we
adopt here, for simplicity, the majority rule. When an opinion is
shared by two interdependent individuals, such as a married couple,
because their social relationship is strong and close, they will tend
to maintain their opinion against outside influence. If their opinions
differ initially, they tend to eventually resolve their differences
and share the same opinion. In the process of resolving their
differences, however, they can be significantly influenced by outside
forces, e.g., mass media, and thus we assume that they often end up
sharing the majority opinion. When we increase the number of
interdependent links between the coupled networks, the transition
changes from a pure second order phase transition to a hybrid phase
transition and finally to a seemingly abrupt transition. The hybrid
transition contains both a second order and an abrupt transition. The
model type of the NCO model on coupled networks also changes as the
number of interdependent links increases, and thus the system goes
from being a kind of nonconsensus opinion model to being a kind of
consensus opinion model. This suggests that strong interactions
between different social groups is pushing our world in the direction
of becoming more uniform in their opinions.

The paper is organized as follows, in Sec.~\ref{percolation} we
revisit some important concepts on the topology of opinion clusters
and percolation. We then present the results and discussions on NCO
and NCO$W$ model in Sec.~\ref{NCO}, on ICO model in Sec.~\ref{ICO} and
on NCO model on coupled networks in Sec.~\ref{NCOCOUPLE}. Finally, we
present our summary in Sec.~\ref{conclusion}.

\section{Topology of Opinion Clusters and Percolation}\label{percolation}

In recent decades, many researchers have studied how network topology
affects the processes that evolve in them. Examples of such processes
are the spreading of rumors, opinions, diseases, and percolation
\cite{Complexnetwork, Boc_01,Dor_02,Pas_01, Vespignaniprl2001,
  LewiStone, KitsakNatureP, Cohenprl2003, Braunsteinprl2003,
  quenched_shlomo, percolation}. Classical percolation processes deal
with the random failure of nodes (or links) and present a geometrical
second order phase transition with a control parameter $p$ that
represents the fraction of nodes (or links) remaining after a random
failure of a fraction $1-p$ of nodes (or links).

There exists a critical probability $p_c$ above which a ``giant
component'' (GC) appears. The number of nodes in the GC, $S_1$, is
called the order parameter of the phase transition. Below criticality
there is no GC and only finite clusters exist. For $p<p_c$ the size
distribution of the clusters is $n_s \sim s^{- \tau}$ with a cutoff
that diverges when approaching $p_c$. At criticality, in the
thermodynamic limit, the size of the second largest component $S_2$
diverges at $p_c$ as $S_2 \sim |p-p_c|^{- \gamma}$ just as the
susceptibility with the distance to the critical temperature. For
large networks ($N \to \infty$), $p_c = 1/(\kappa-1)$, where $\kappa$
is the branching factor given by $\kappa= \langle k^2 \rangle /\langle
k \rangle$, where $\langle k \rangle$ and $\langle k^2 \rangle$ are
the first and second moments of the degree distribution $P(k)$ of the
network respectively \cite{Cohen}. We perform all our simulations on
both Erd\"{o}s-R\'{e}nyi (ER) networks \cite{ER1, ER1959, Bollobas}
and scale-free (SF) networks \cite{SF1}.  ER networks are
characterized by a Poisson degree distribution, $P(k)=e^{- \langle k
  \rangle} \langle k \rangle^k /k!$. In SF networks the degree
distribution is given by a power law, $P(k)\sim k^{-\lambda}$, for
$k_{\rm min}\leq k \leq k_{\rm max}$, where $k_{\rm min}$ is the
lowest degree of the network and $k_{\rm max}$ is the highest degree
of the network. For random SF networks $k_{\rm max} \sim
N^{1/(\lambda-1)}$ is the degree cutoff \cite{Cohen}, where $N$ is the
system size and $\lambda$ is the broadness of the distribution.

We begin by examining percolation in ER networks. At criticality,
percolation in ER networks is equivalent to percolation on a Cayley tree
or percolation at the upper critical dimension $d_c=6$ where all the
exponents have mean field (MF) values with $\tau=5/2$ and $\gamma=1$.
Note that in the ER case, the mass of the incipient infinite cluster
$S_1$ scales as $N^{2/3}$ at criticality. We can understand this result
by using the framework of percolation theory for the upper critical
dimension $d_c=6$. Since $S_1\sim R^{d_f}$ and $N\sim
R^d$ (where $d$ is the dimension of the initial lattice, $d_f$ the
fractal dimension, and $R$ the spatial diameter of the cluster), it
follows that $S_1\sim N^{d_f/d_c}$ and since $d_c=6$ and $d_f=4$ we obtain
$S_1\sim N^{2/3}$ \cite{Braunsteinprl2003}.

For SF networks, the GC at criticality is $S_1\sim N^{2/3}$ for
$\lambda>4$, and $S_1\sim N^{(\lambda-2)/(\lambda-1)}$ for
$3<\lambda\leq 4$ \cite{cohenpre2002}. For SF networks, with $\lambda <
3$, $\langle k^2 \rangle \to \infty$ when $N \to \infty$ because $k_{\rm
  max} \to \infty$ and thus $p_c=0$, making these networks extremely
robust against random failures \cite {Cohen}.

However if we decrease $k_{\rm max}$ by targeting and removing the
highest degree nodes (hubs), $p_c$ is finite \cite{cohenprl2001} and we recover a
second order phase transition with MF exponents as for ER networks. We
will show below that this is also true for our model here.  A similar
MF behavior in SF networks with $\lambda < 3$ was found also by Valdez
{\it et al.} \cite{L.Valdez} for the percolation of susceptible
clusters during the spread of an epidemic.  In our simulations we
always choose $k_{\rm min} = 2$ for SF networks in order to ensure
that they are almost fully connected \cite{Cohen}.

\section{The NCO Model}\label{NCO}

In the NCO model \cite{NCO} on a single network with $N$ nodes, opinion
$\sigma_+$ and $\sigma_-$ are initially randomly assigned to each node
with a fraction of $f$ and $1-f$ respectively. The basic assumption of
the NCO model is that the opinion of an agent is influenced by both its
own opinion and the opinions of its nearest neighbors (the agent's
friends). The opinion formation rule states that at each time step, each
node adopts the majority opinion, which includes {\it both\/} the
opinions of its neighbors and itself. If there is a tie, the node's
opinion will remain unchanged. Using this rule, each node is tested at
each simulation step to see whether its opinion has changed. All these
updates are performed simultaneously and in parallel until no more
changes occur and a steady state is reached.

\begin{figure}[ht]
\includegraphics[width=8cm,height=8cm]{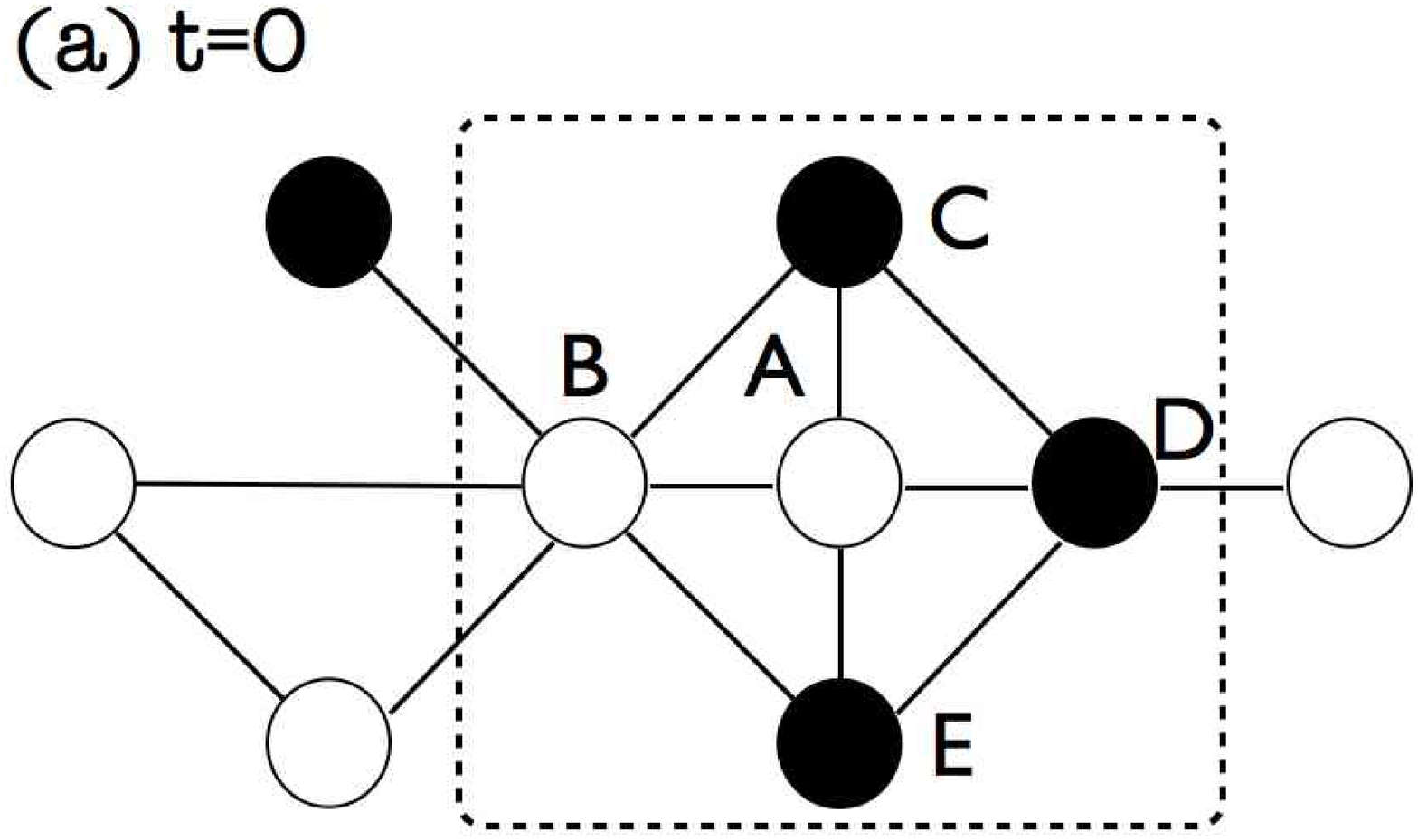}
\includegraphics[width=8cm,height=8cm]{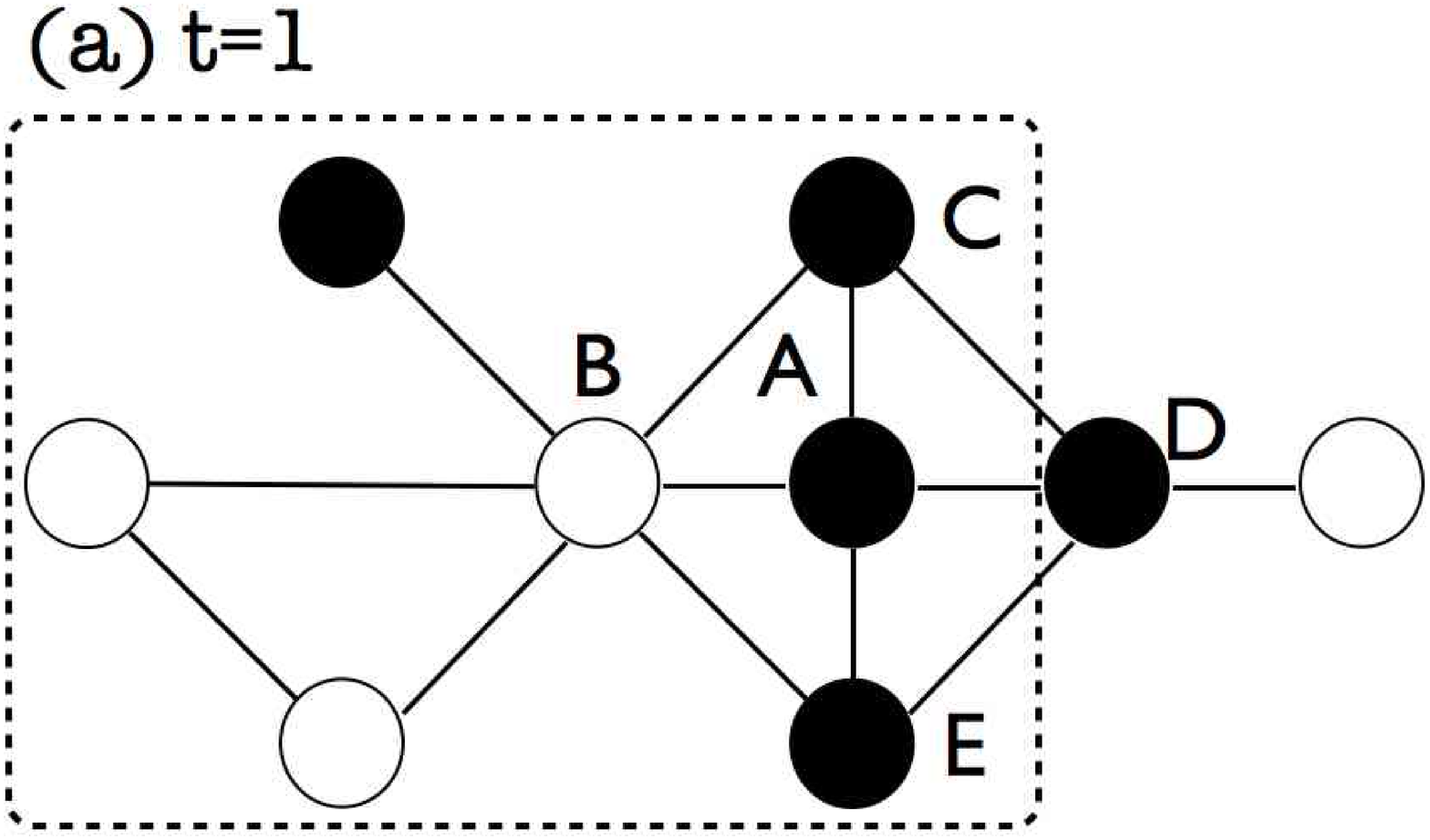}
\includegraphics[width=8cm,height=8cm]{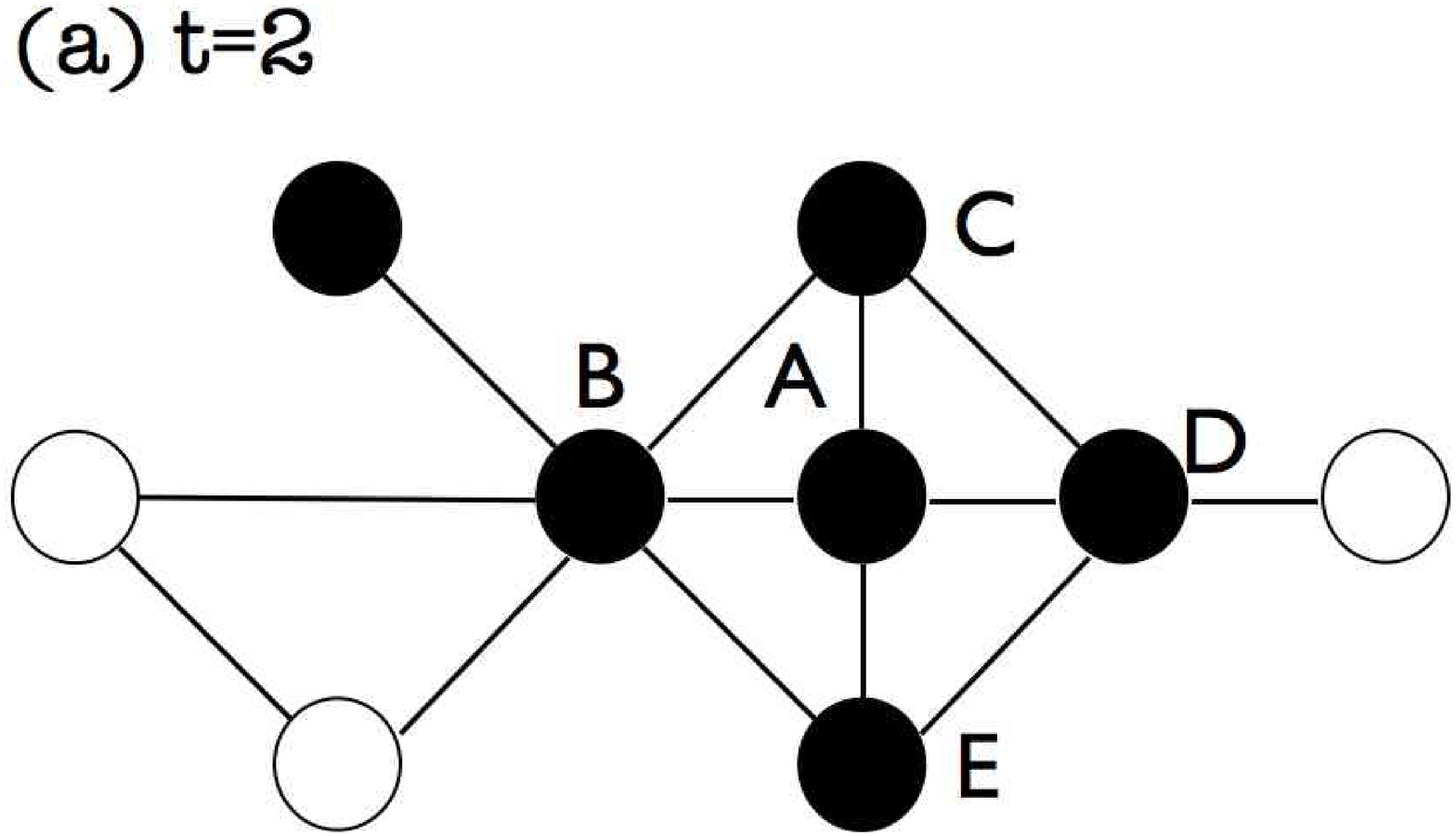}
\caption{Dynamics of the NCO model showing the approach to a stable
  state on a network with $N=9$ nodes. (a) At t=0, five nodes are
  randomly assigned to be $\sigma_+$ (empty circle), and the remaining
  four nodes are assigned with $\sigma_-$ (solid circle). In the set
  comprising of node $A$ and its 4 neighbors (dashed box), node $A$ is
  in a local minority opinion, while the remaining nodes are not. Thus
  at the end of this simulation step, node $A$ is converted into
  $\sigma_-$ opinion. (b) At t=1, in the set of nodes comprising node
  $B$ and its 6 neighbors (dashed box), node $B$ becomes in a local
  minority opinion, while the remaining nodes are not. Thus, node $B$ is
  converted into $\sigma_-$ at the end of simulation step $t=1$. (c) At,
  $t=2$, the nine nodes system reaches a stable
  state. \label{fig:NCOf1}}
\end{figure}

Figure~\ref{fig:NCOf1} demonstrates the dynamic behavior of the NCO
model on a small network with nine nodes. At time $t=0$, five nodes
are randomly assigned opinion $\sigma_+$ (empty circle), and the
remaining four, opinion $\sigma_-$ (solid circle). After checking the
status of each node, we find that only node A belongs to a local
minority with opinion $\sigma_+$, so at the end of this time step,
node A changes its opinion to $\sigma_-$. At time t=1 only node B
belongs to a local minority, so at the end of this time step, the
opinion of node B will be updated to $\sigma_-$. At time t=3, every
node has the same opinion as its local majority, where the final
nonconsensus steady state is reached.

\subsection{Simulation Results}

\begin{figure}[ht]
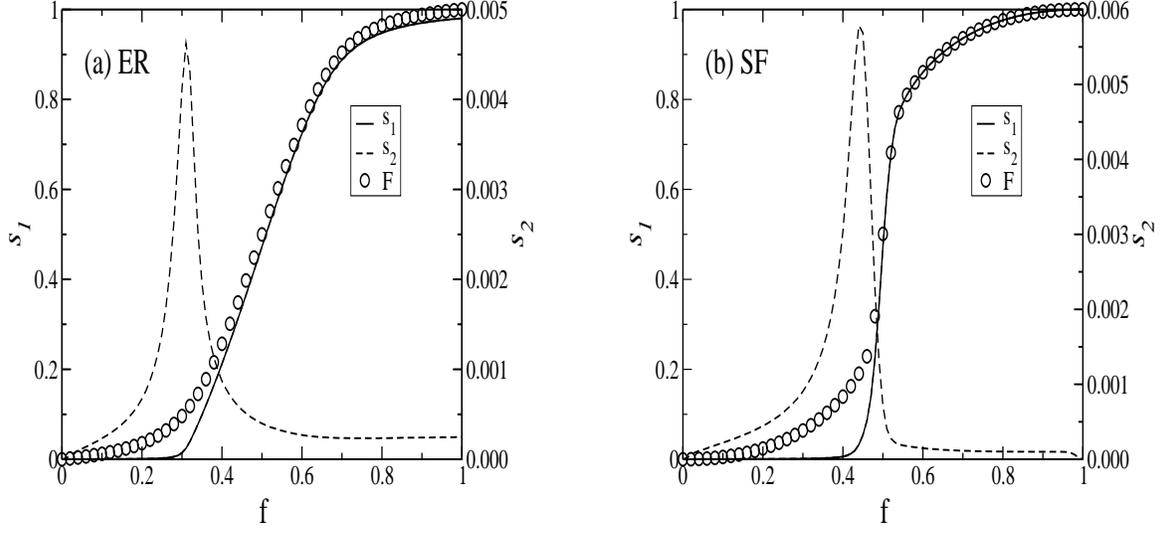

\includegraphics[width=7cm,height=7cm]{NCO_fig1a.eps}
\hspace{1cm}
\includegraphics[width=7cm,height=7cm]{NCO_fig1b.eps}\\
\vspace{2cm}
\caption{Plots of $s_1$, $s_2$ and $F$ of opinion $\sigma_+$ as a
  function of $f$ with network size $N=10000$, for (a) ER networks with
  $\langle k\rangle=4$ and (b) SF networks with $\lambda=2.5$ and
  $k_{\rm min}=2$. All simulations were done for $10^4$ networks
  realizations.\label{fig:NCOf2}}
\end{figure}

In the steady state $s_1=S_1/N$ is the normalized size of the largest
opinion $\sigma_+$ cluster, $s_2=S_2/N$ is the normalized size of the
second largest opinion $\sigma_+$ cluster, and $F$ is the normalized
fraction of opinion $\sigma_+$ nodes. Figure~\ref{fig:NCOf2} shows
plots of $s_1$, $s_2$, and $F$ as a function of the initial fraction
$f$ of the opinion $\sigma_+$ nodes for both ER and SF networks. We
find that due to the symmetrical status of both opinions, $F$ is a
monotonically increasing function of $f$ with symmetry around
$f=0.5$. Figure~\ref{fig:NCOf2} also shows the emergence of a second
order phase transition. Note that there is a critical threshold $f_c$,
which is characterized by the sharp peak of $s_2$. Below $f_c$, $s_1$
approaches zero, where only the majority opinion can form steady
clusters, and above $f_c$, $s_1$ increases as $f$ increases and a
state with stable coexistence of both majority and minority opinion
clusters appears. Although for both ER and SF networks $f_c<0.5$ as
expected, ER networks have smaller values of $f_c$ than SF networks
for the same average degree $\langle k\rangle$. For example for SF
networks with $k_{\rm min}=2$ and $\lambda=2.5$ where $\langle
k\rangle \approx 5.5$, $f_c \approx 0.45$, while for the same average
degree $\langle k\rangle = 5.5$, for ER networks, $f_c \approx
0.4$. These differences in $f_c$ indicate that the minorities in SF
networks need more initial supports to form final steady state
clusters, than minorities in ER networks. This might be understood due
to the high degree nodes (hubs) of a SF network. In the NCO model a
hub, because of its large number of connections, is strongly
influenced by its neighbors which are with high probability of the
majority opinion, and puts the minority opinion at a
disadvantage. Reference~\cite{NCO} presents also studies of the NCO
model in a two-dimensional Euclidean lattice and of the NCO model on
real-world networks.

\begin{figure}[ht]
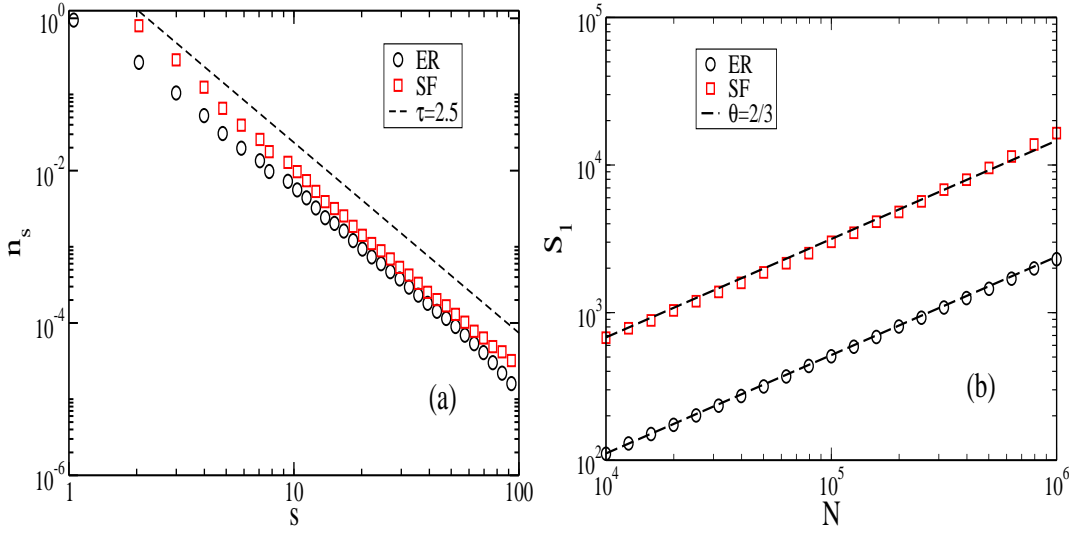

\includegraphics[width=7cm,height=7cm]{NCO_fig2.eps}
\includegraphics[width=7cm,height=7cm]{S1_N.eps}\\
\vspace{0.5cm}
\caption{ (a) Plots of $n_s$ as a function of $s$ at criticality for
  ER networks with $\langle k\rangle=4$ and SF networks with
  $\lambda=2.5$ and $k_{\rm min}=2$. The dashed line is a guide to
  show that the slope obtained is $\tau=2.5$. (b) Plots of $S_1$ as a
  function of $N$ at criticality for ER networks with $\langle
  k\rangle=4$ and SF networks with $\lambda=2.5$ and $k_{\rm
    min}=2$. The dashed lines are guides to show that the slope
  obtained is $\theta \approx 2/3$. All simulations were done for
  $N=10000$ and over $10^4$ networks realizations.\label{fig:NCOf3}}
\end{figure}

We next present numerical simulations indicating that the phase
transition observed in the NCO model is in the same universality class
as regular MF percolation. Percolation in random networks (e.g., ER
and SF networks with $\lambda > 4$ for random failures or all
$\lambda$ for targeted attacks) \cite{percolation,quenched_shlomo,
  Complexnetwork} is obtained by MF theory, which predicts that at
criticality the cluster size distribution is $n_s \sim s^{-\tau}$ with
$\tau=2.5$ and $S_1 \sim N^{\theta}$, where $\theta=d_f/d_c$ with
$d_f=4$ and $d_c=6$ represent the fractal and the upper critical
dimension of percolation respectively and thus $\theta=2/3$ (See
Sec.~\ref{percolation}).  Figure~\ref{fig:NCOf3}(a) shows the finite
cluster size distribution $n_s$ of the $\sigma_+$ opinion cluster as a
function of $s$ at criticality ($f=f_c$). Figure~\ref{fig:NCOf3}(b)
show $S_1$ at criticality $f_c$ as a function of $N$ for ER and SF networks
with $\langle k \rangle=4$ and $\lambda=2.5$, respectively.  Note that
in both networks $\tau \approx 2.5$ and $\theta \approx 2/3$. These
two exponents strongly indicate that the NCO model in random networks
behaves like a second order phase transition that belongs to the same
universality class as regular MF percolation.

\begin{figure}[ht]
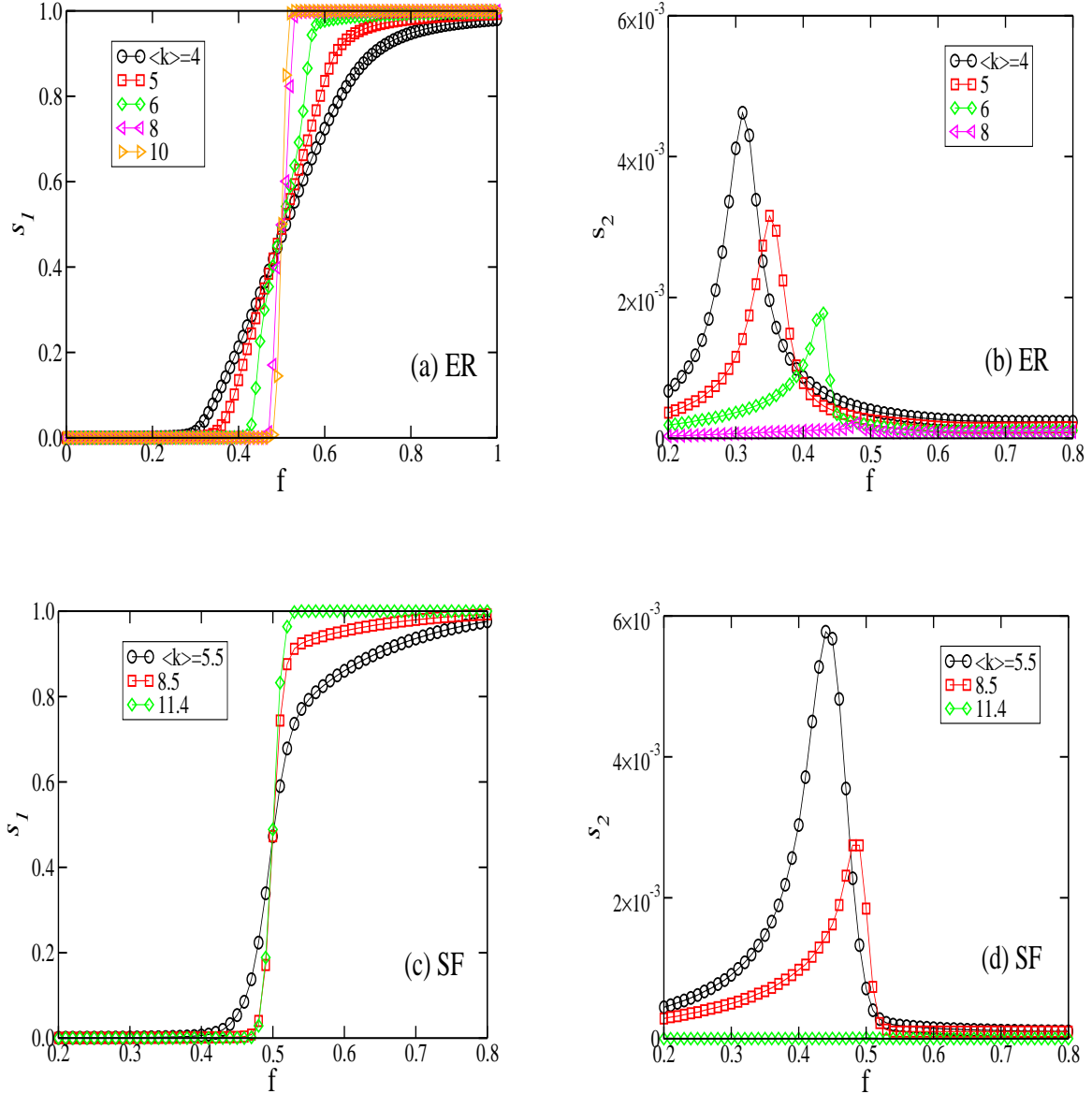

\includegraphics[width=7cm,height=7cm]{NCO_fig3a.eps}
\hspace{1cm}
\includegraphics[width=7cm,height=7cm]{NCO_fig3b.eps}\\
\vspace{1.5cm}
\includegraphics[width=7cm,height=7cm]{NCO_fig3c.eps}
\hspace{1cm}
\includegraphics[width=7cm,height=7cm]{NCO_fig3d.eps}
\vspace{0.5cm}
\caption{Plots of (a) $s_1$ and (b) $s_2$ of opinion $\sigma_+$ as a
  function of $f$ for ER networks with different values of $\langle
  k\rangle$ for $N=10000$. (c) $s_1$ and (d) $s_2$ of opinion
  $\sigma_+$ as a function of $f$ for SF networks with different
  values of $\langle k\rangle$ for $N=10000$ and $\lambda=2.5$. The
  solid lines are guides for the eyes. All simulations were done for
  $10^4$ networks realizations.
 \label{fig:NCOf4}}
\end{figure}

In our above results we focus on networks that have a relatively low
average degree $\langle k \rangle$. We test the model for networks
with higher average degrees. In SF networks we increase $\langle k
\rangle$ by increasing the value of $k_{\rm
  min}$. Figure~\ref{fig:NCOf4} shows $s_1$ and $s_2$ as a function of
$f$ for different values of $\langle k \rangle$ for ER networks and SF
networks. As the values of $\langle k \rangle$ increase, i.e., as the
network becomes increasingly condensed and the number of interactions
between agents increases, a sharper change of $s_1$ at a critical
threshold is observed. This may suggest (but can not be proved by
simulations) the existence of a critical value $\langle k \rangle=k_c$
that is strongly affected by the topology of the network. Below $k_c$,
as $\langle k \rangle$ increases, $f_c$ shifts to the right, as can be
seen from the shift of the peak of $s_2$. Above $f_c$ two opinions can
continue to coexist and remain stable. Above $k_c$ the smooth second
order phase transition is replaced by a sharp jump of $s_1$ at
approximately $f=0.5$ that is accompanied by the disappearance of the
peaks of $s_2$. Note also that as the values of $\langle k \rangle$
increase, the region in which two opinions coexist becomes
increasingly smaller and approach zero for very large values of
$\langle k \rangle$ possibly above $k_c$. In terms of the NCO model,
as the number of connections between individuals increase, the opinion
of each individual becomes less important and each individual becomes
increasingly susceptible to the influence of the majority opinion
across the entire system.  Thus the majority opinion can easily
overwhelm the minority opinion, causing the critical behavior of the
NCO model, the second-order phase transition, to disappear at large
$\langle k \rangle$ and the NCO model to converge to the majority
voter model yielding a possible global consensus throughout the
system. Note that analytical arguments for an abrupt transition of the
NCO model at large $\langle k \rangle$ are given in
Ref.~\cite{Percolate_or_die}.

\begin{figure}[ht]
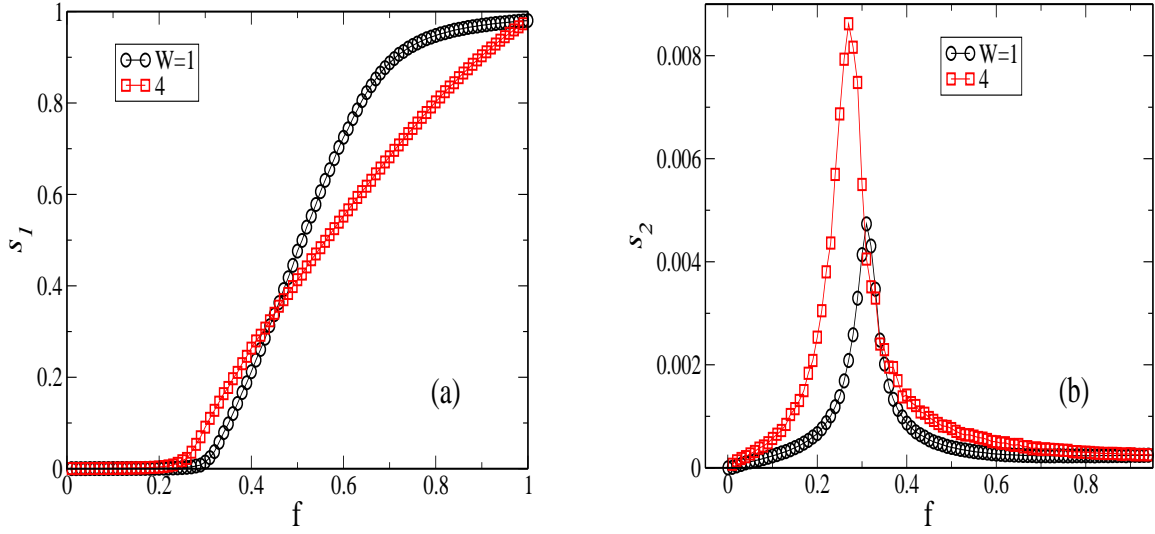

\includegraphics[width=7cm,height=7cm]{NCOW_S1.eps}
\hspace{1cm}
\includegraphics[width=7cm,height=7cm]{NCOW_S2.eps}\\
\vspace{0.2cm}
\caption{Plots of (a) $s_1$ and (b) $s_2$ of opinion $\sigma_+$ as a
  function of $f$, for ER networks with different values of $W$ for
  $\langle k\rangle=4$. The solid lines are guides for the eyes. All
  simulations were done with $N=10000$ and for $10^4$ networks
  realizations.\label{fig:NCOf5}}
\end{figure}

How can one help the minority opinions to survive? As we have seen, as
the number of friends of an agent increases, the importance of the
agent's own opinion decreases. In this way the majority opinion
gradually eliminates the minority opinion. If we generalize the NCO
model by adding a weight value $W$ to each agent's own opinion, as $W$
of an agent increases, the influence of the opinion of the agent's
neighbors decreases. We call this generalization of the NCO model the
NCO$W$ model. As in the NCO model, in the NCO$W$ model we change an
agent's opinion if he is in a local minority but we also weight the
agent's own opinion $W$ times more than its nearest neighbors. The NCO
model is actually a special case of the NCO$W$ model in which $W=1$.
Figure~\ref{fig:NCOf5} shows plots of $s_1$ and $s_2$ as a function of
$f$ for both $W=1$ and $W=4$. Note that as $W$ increases, the
second-order phase transition becomes flatter and the peak of the
$s_2$ shifts to the left, which indicates a smaller critical threshold
$f_c$. The smaller value of $f_c$ for larger values of $W$ means that
from the minority point of view, it needs fewer initial supporters to
form and maintain stable finite clusters.  When weight is added to the
agents own opinion (indicating stubbornness) they become less
susceptible to outside influence. Thus in the NCO$W$ model the
majority is aided when the agents make more friends, but the minority
in turn is aided when the agents treat their own opinion as more
important than their friends' opinions.

\section{The ICO Model}\label{ICO}

\begin{figure}[ht]
\includegraphics[width=6cm,height=6cm]{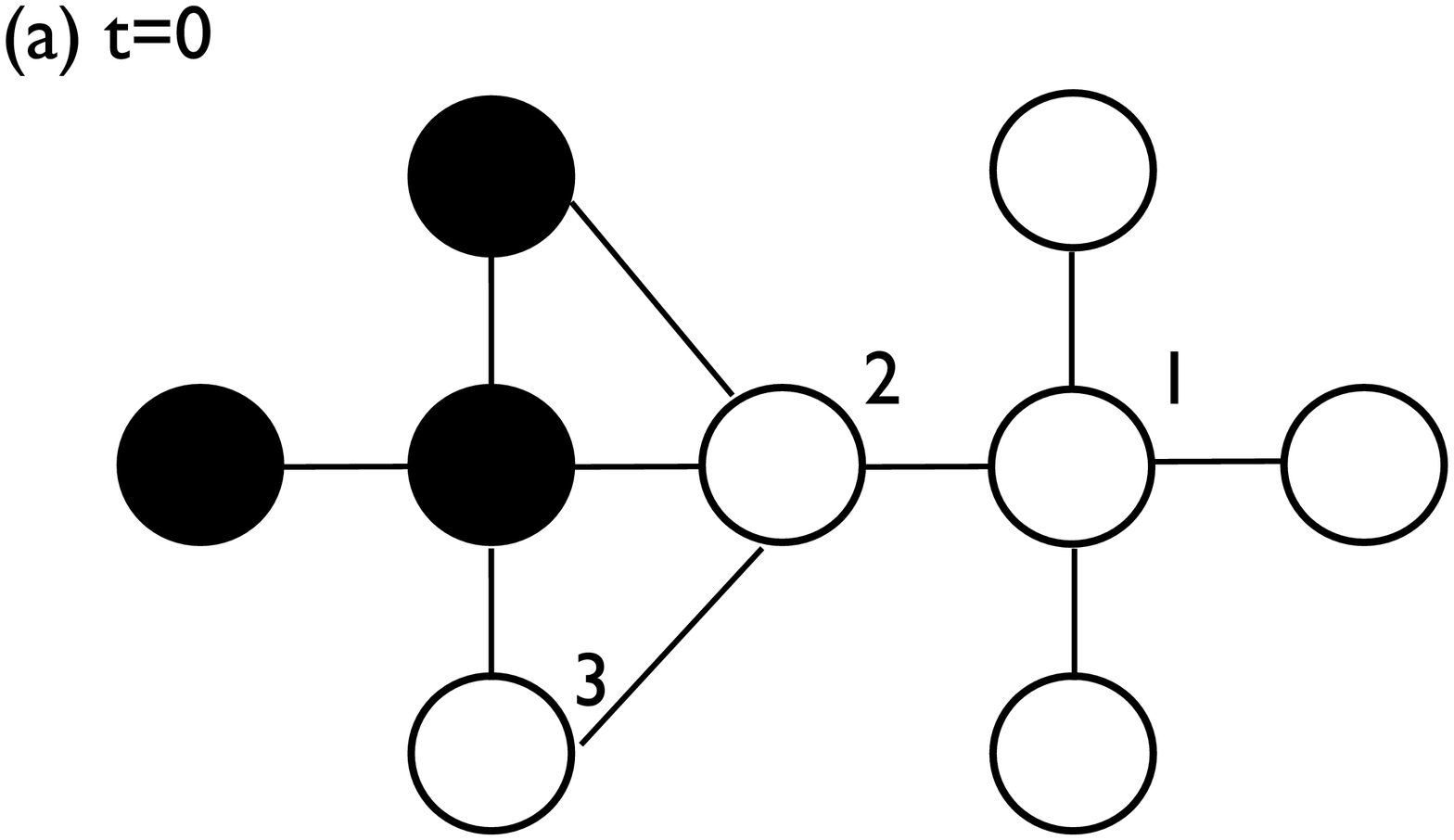}
\hspace{1cm}
\includegraphics[width=6cm,height=6cm]{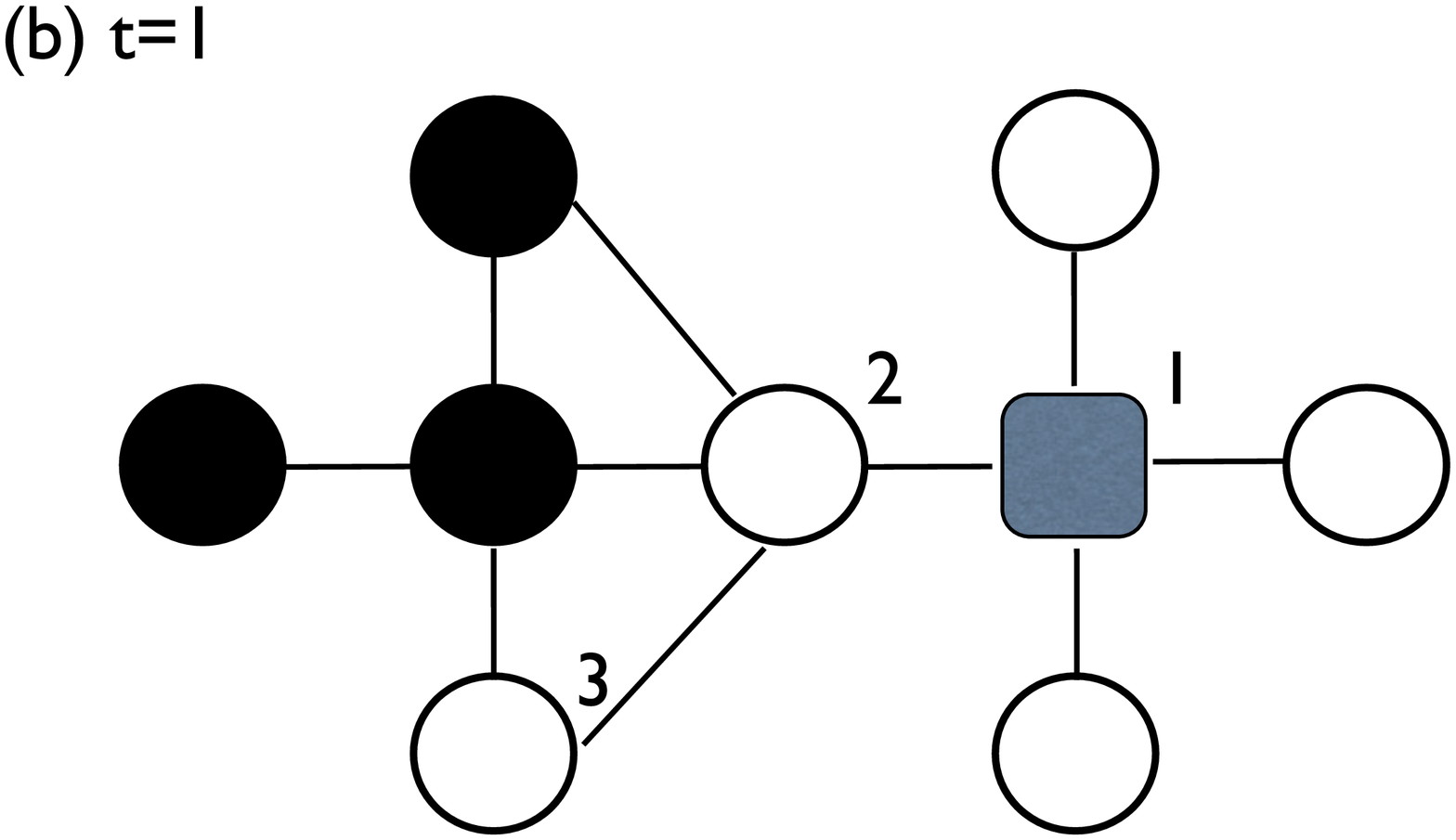}\\
\vspace{3cm}
\includegraphics[width=6cm,height=6cm]{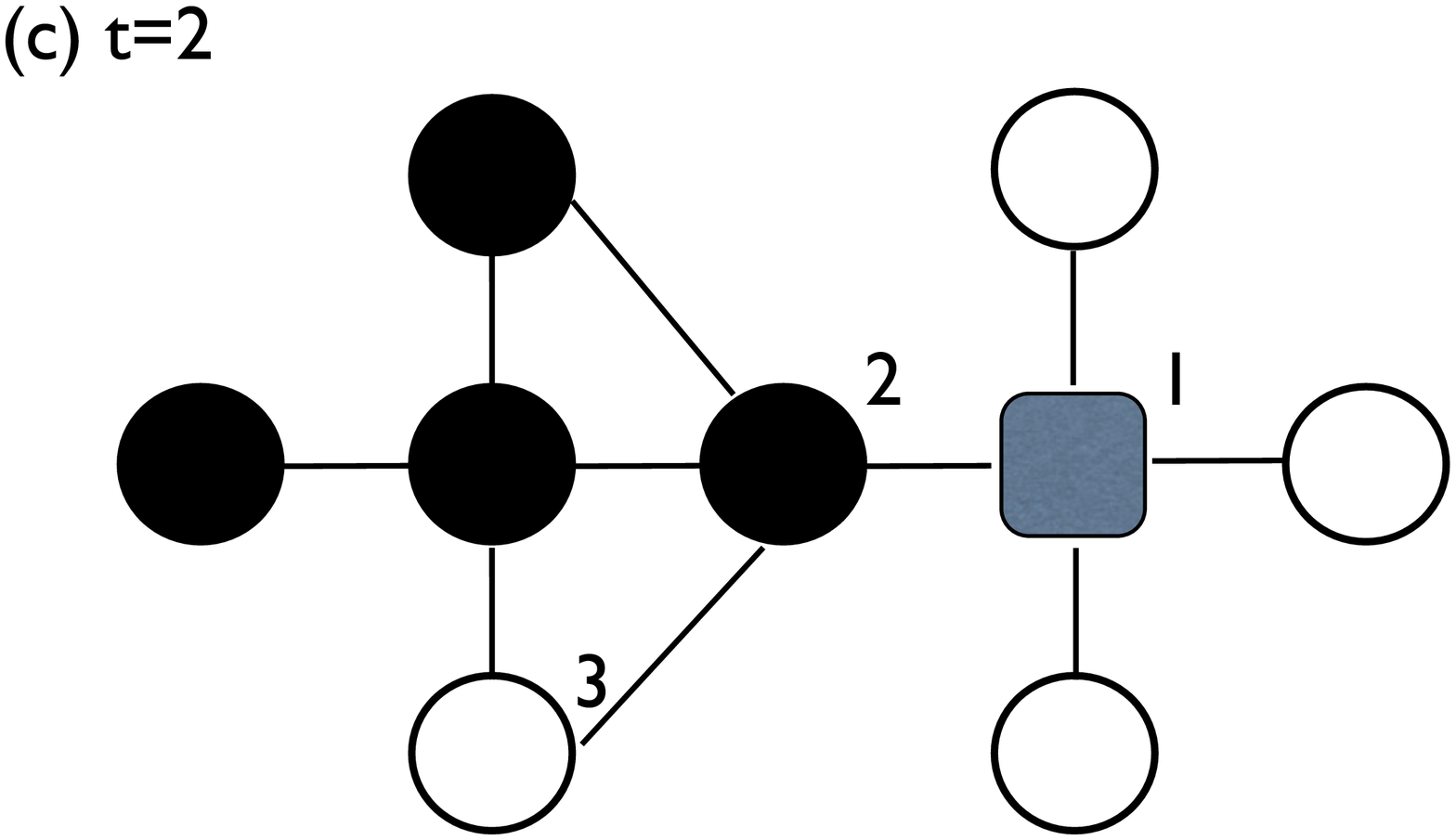}
\hspace{1cm}
\includegraphics[width=6cm,height=6cm]{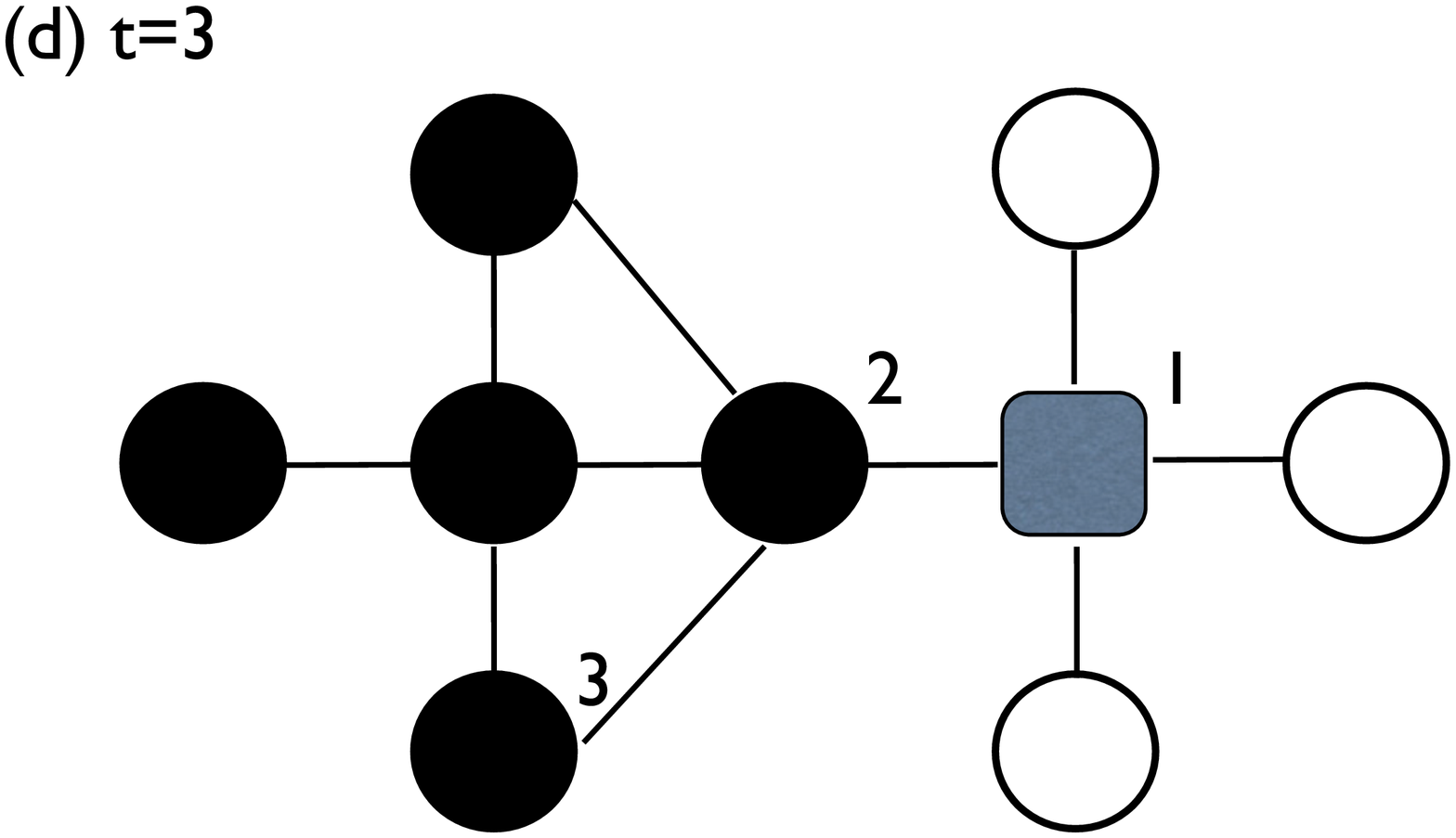}\\
\vspace{1cm}
\caption{Schematic plot of the dynamics of the ICO model showing the
  approach to a stable state on a network with $N=9$ nodes. (a) At
  $t=0$, we have a stable state where opinion $\sigma_+$ (open circle)
  and opinion $\sigma_-$ (filled circle) coexist. (b) At $t=1$, we
  change node $1$ into a inflexible contrarian (filled square), which
  will hold $\sigma_-$ opinion. Node $2$ is now in a local minority
  opinion while the remaining nodes are not. Notice that node $1$ is an
  inflexible contrarian and even if he is in the local minority he will
  not change his opinion. At the end of this simulation step, node $2$
  is converted into $\sigma_-$ opinion. (c) At $t=2$, node $3$ is in a
  local minority opinion and therefore will be converted into $\sigma_-$
  opinion. (d) At $t=3$, the system reaches a stable state where the
  system breaks into four disconnected clusters, one of them composed of
  six $\sigma_-$ nodes and the other three with one $\sigma_+$
  node. \label{fig:ICOf1}}
\end{figure}

The initial configuration of the inflexible contrarian opinion (ICO)
model corresponds to the final steady state of the NCO model in which
two opinions $\sigma_+$ and $\sigma_-$ coexist. At $t=0$ a fraction
$\phi$ of inflexible contrarians of opinion $\sigma_-$ are introduced
into clusters of $\sigma_+$ by replacing nodes of $\sigma_+$. The
inflexible contrarians are agents that hold a strong and unchangeable
$\sigma_-$ opinion, that theoretically could influence the $\sigma_+$
opinion of their neighbors as the system evolves with NCO
dynamics. Because the opinion held by the inflexible contrarians is
unchanging, they function as a quenched noise in the network. The
system evolves according to NCO dynamics until a new steady state is
reached. In this steady state the agents form clusters of two
different opinions above a new threshold $f_c \equiv
f_c(\phi)$. Because the contrarians hold the $\sigma_-$ opinion, the
size of the $\sigma_+$ clusters decreases as $\phi$
increases. Figure~\ref{fig:ICOf1} demonstrates the dynamic of
the ICO model. We use two different strategies to introduce a fraction
$\phi$ of inflexible contrarians. In strategy I we chose the fraction
$\phi$ of nodes with $\sigma_+$ opinion at random. In strategy II the
inflexible contrarians are chosen from the agents with $\sigma_+$
opinion in decreasing order of their connectivity. Strategy II is thus
a targeted strategy.

\subsection{Simulation Results}

\begin{figure}[ht]
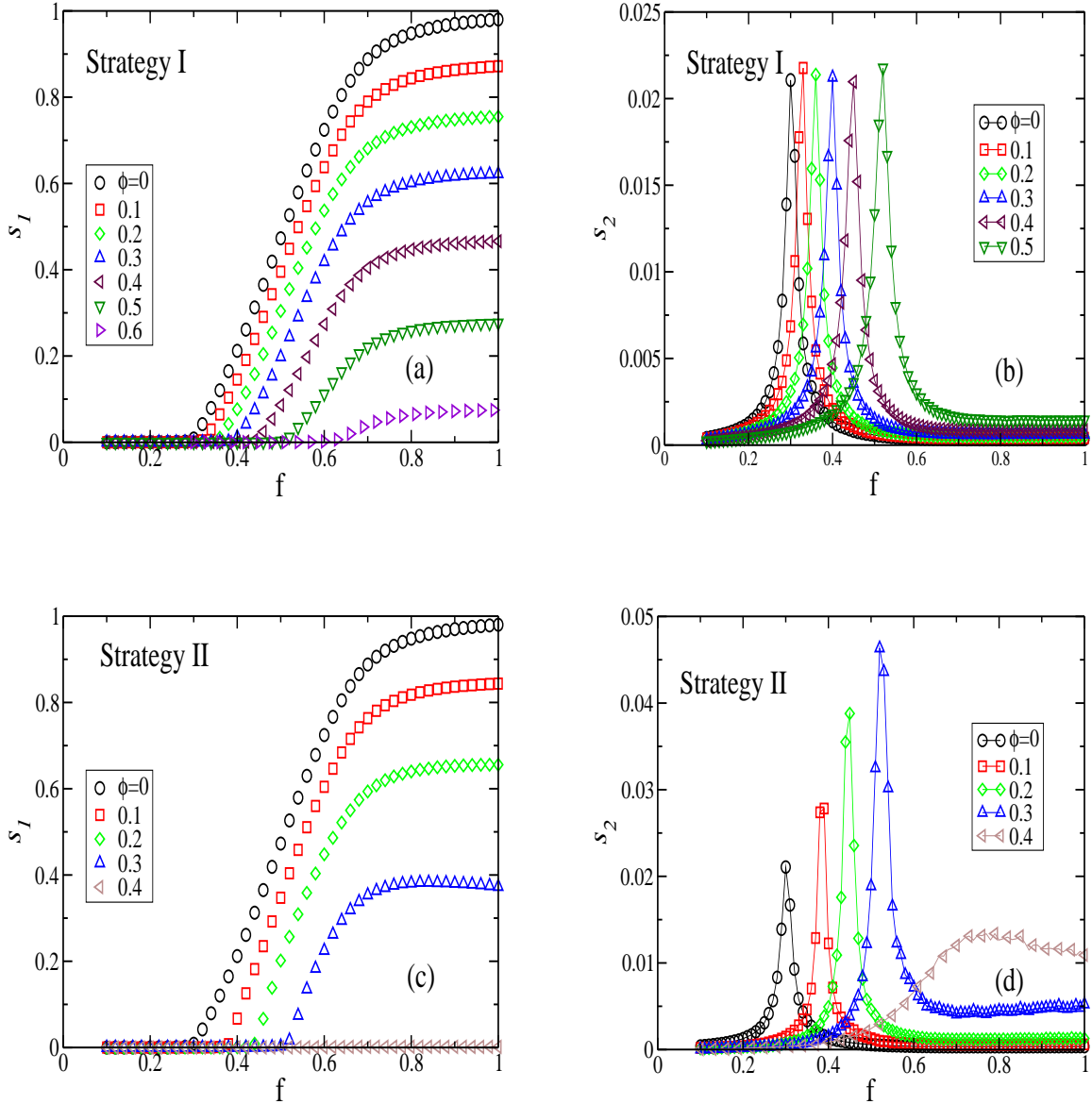

\includegraphics[width=7cm,height=7cm]{ICO_fig1_a.eps}
\hspace{1cm}
\includegraphics[width=7cm,height=7cm]{ICO_fig1_b.eps}\\
\vspace{1.5cm}
\includegraphics[width=7cm,height=7cm]{ICO_fig1_c.eps}
\hspace{1cm}
\includegraphics[width=7cm,height=7cm]{ICO_fig1_d.eps}
\caption{For ER networks with $\langle k\rangle=4$, plots of (a) $s_1$
  and (b) $s_2$ as a function of $f$ for different values of $\phi$
  under strategy I. Plots of (c) $s_1$ and (d) $s_2$ as a function for
  $f$ for different values of $\phi$ under strategy II. The solid
  lines in $s_2$ are guides for the eyes. All simulations were done
  with $N=10000$ and $10^4$ network realizations. \label{fig:ICOf2}}
\end{figure}

\begin{figure}[ht]
\includegraphics[width=8cm,height=8cm]{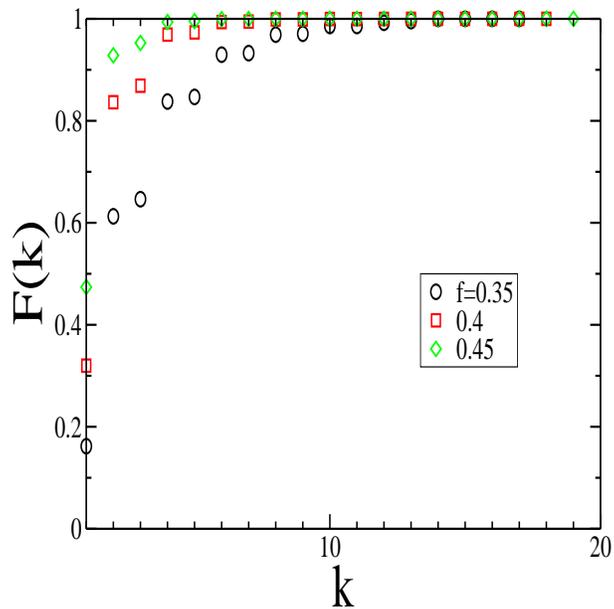}
\caption{$F(k)$ as a function of $k$ for different values of $f$ for ER
  networks with $\langle k\rangle=4$ . We can see that as $f$ increases
  $F(k) \to 1$ for smaller values of $k$. All simulations were done
  with $N=10000$ and $10^4$ network realizations. \label{fig:ICOf3}}
\end{figure}

We present our simulation results for ER networks with $\langle k
\rangle=4$ and $N=10^5$. For simulation results of ICO model for SF
networks see Ref.~\cite{ICO}. Figure~\ref{fig:ICOf2} shows plots of
$s_1$ and $s_2$ as a function of $f$ for different values of $\phi$
for strategies I and II, respectively. Note that the ICO model
inherits some of the properties of the NCO model. This is the case
because there is a smooth phase transition with a critical threshold
$f_c$, where $f_c$ is characterized by the sharp peak of
$s_2$. However, for the ICO model, $f_c$ is also a function of
$\phi$. Thus we denote the new $f_c$ in the ICO model by
$f_c(\phi)$. We find that, as $\phi$ increases, the critical value
$f_c(\phi)$ increases, which means that the largest cluster composed
of $\sigma_+$ agents becomes less robust due to the increase in the
number of inflexible contrarians of opinion $\sigma_-$. Note also that
for $f>f_c(\phi)$, $s_1$ decreases as $\phi$ increases. Thus, we
conclude that inflexible contrarians with opinion $\sigma_-$ have two
effects: (i) they increase the value of $f_c(\phi)$ and thus the
$\sigma_+$ opinion needs more initial support in order to survive, and
(ii) they decrease the size of the largest $\sigma_+$ opinion cluster
at $f>f_c(\phi)$. Note also that in the ICO model when $\phi$ is large
the largest $\sigma_+$ cluster is fully destroyed and the second-order
phase transition is lost. This is probably due to the fact that when
$\phi$ is large, minority groups do not have high degree nodes and
thus their average connectivity becomes smaller than $1$ and, as a
consequence, will no longer be able to form stable clusters
\cite{ER1}. As expected (see Fig.~\ref{fig:ICOf2}) strategy II is more
efficient in destroying the largest minority component. This is
plausible because, when selecting the initial fraction $\phi$ of
inflexible contrarians using a targeted strategy, almost all the
inflexible contrarians will be in the largest initial $\sigma_+$
cluster since this cluster includes most of the high degree
nodes. Figure~\ref{fig:ICOf3} test this hypothesis and shows at the
final stage of the NCO the ratio $F(k)$, which is the number of nodes
within the GC of $\sigma_+$ opinion with degree $k$ divided by the
total number of nodes of opinion $\sigma_+$ with degree $k$ in the
entire network system, for different values of $f$. We find that for
large values of $k$, $F(k) \to 1$. These results support our previous
hypothesis that almost all the high degree nodes belong to the largest
cluster, and this explains why strategy II is more efficient than
strategy I.

\begin{figure}[ht]
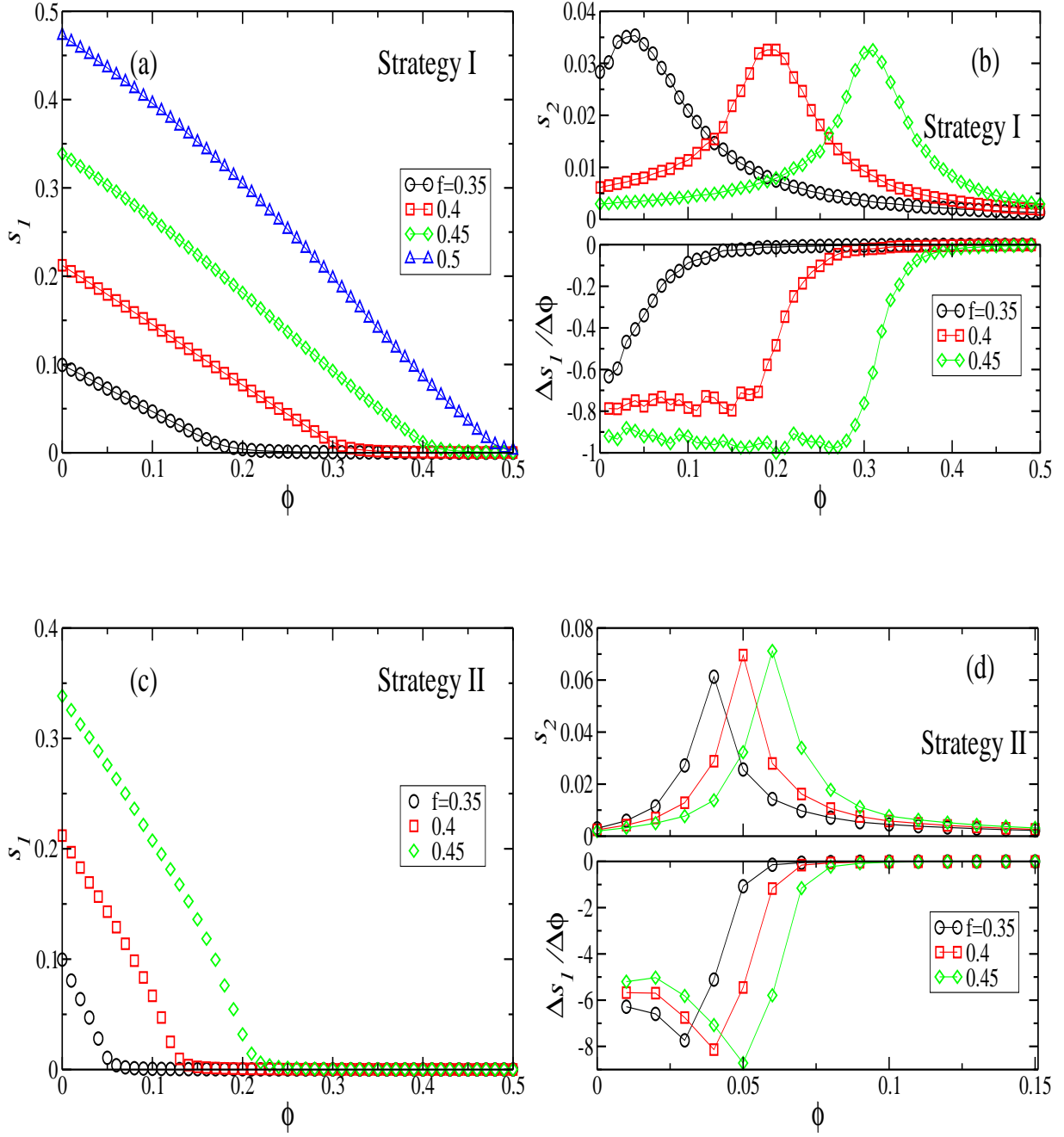

\includegraphics[width=8cm,height=8cm]{ICO_fig2_a.eps}
\includegraphics[width=8cm,height=8cm]{ICO_fig2_b.eps}\\
\vspace{1.5cm}
\includegraphics[width=8cm,height=8cm]{ICO_fig2_c.eps}
\includegraphics[width=8cm,height=8cm]{ICO_fig2_d.eps}
\caption{ For ER networks with $\langle k\rangle=4$, plots of (a)
  $s_1$ (b)(top) $s_2$ and (b)(bottom) $\Delta S_1/\Delta \phi$ as a
  function of $\phi$ for different values of $f$ under strategy
  I. Plot of (c) $s_1$, (d)(top) $s_2$ and (d)(bottom) $\Delta
  S_1/\Delta \phi$ as a function of $\phi$ for different values of $f$
  under strategy II. The solid lines in $s_2$ and $\Delta S_1/\Delta
  \phi$ are guides for the eyes. All simulations were done with
  $N=10000$ and $10^4$ network realizations.\label{fig:ICOf4}}
\end{figure}

\begin{figure}[ht]
\includegraphics[width=14cm,height=14cm]{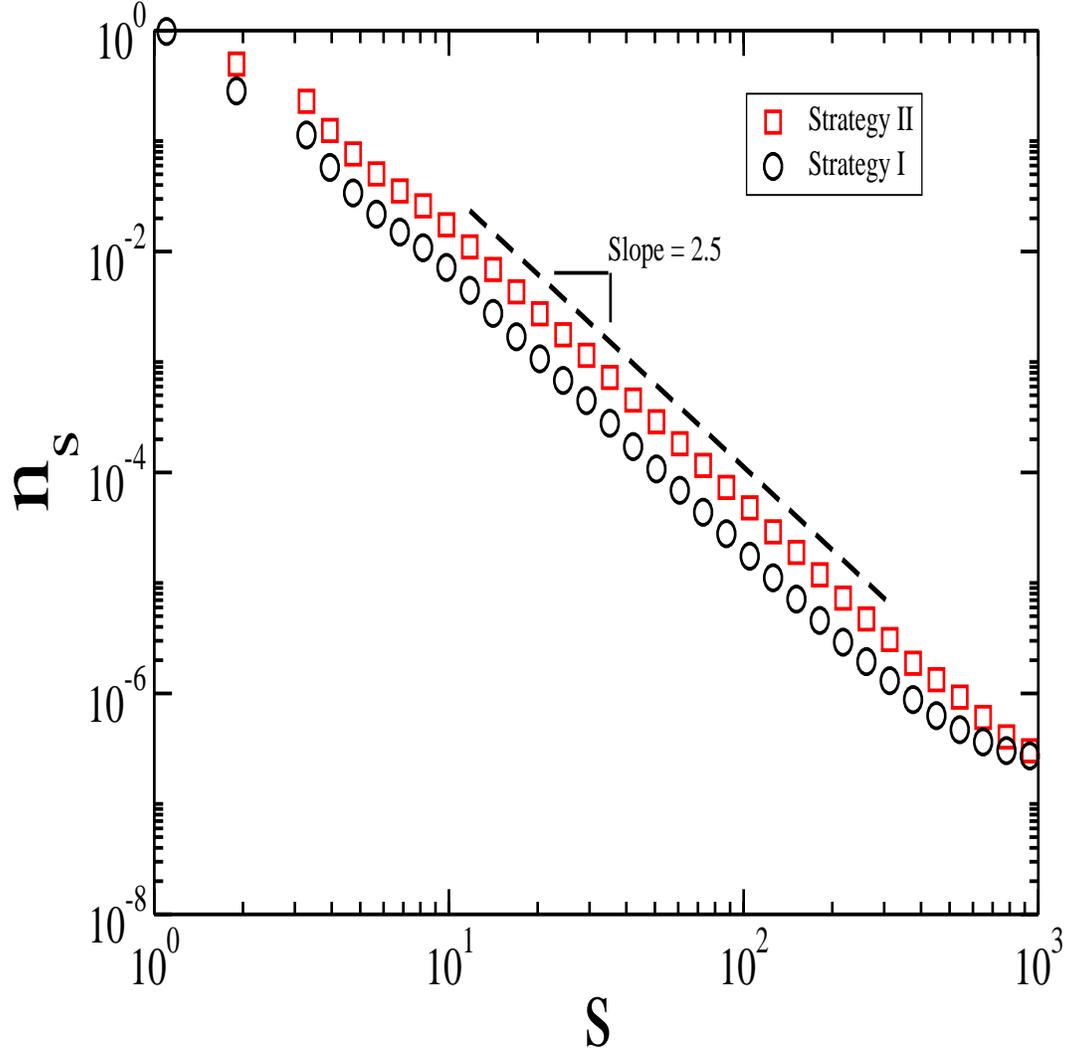}
\caption{Plots of $n_s$ as a function of $s$ for both strategies at
  $f_c(\phi)$. For the random strategy, $f_c(\phi)=0.36$ and for the
  targeted strategy, $f_c(\phi)=0.45$ when $\phi=0.2$. The dashed line
  is a guide to show that the slope obtained is $\tau \approx
  5/2$. All simulations were done with $N=10000$ and $10^4$ network
  realizations. \label{fig:ICOns}}
\end{figure}

We next test whether the ICO model undergoes a phase transition as a
function of $\phi$ and what it its type. Figures~\ref{fig:ICOf4}(a)
and \ref{fig:ICOf4}(c) show plots of $s_1$ as a function of $\phi$ for
different values of $f$ for strategy I and strategy II,
respectively. Figures~\ref{fig:ICOf4}(b)(top) and
\ref{fig:ICOf4}(d)(top) show plots of $s_2$ as a function of $\phi$
for different values of $f$ for strategy I and strategy II,
respectively. We can see that in both strategies $s_2$ has a peak at
$\phi=\phi_c(f)$, which is a characteristic of a second-order phase
transition.  Figures~\ref{fig:ICOf4}(b)(bottom) and
\ref{fig:ICOf4}(d)(bottom) further support the presence of a second
order phase transition by showing plots of the derivative of $s_1$
with respect to $\phi$ for different values of $f$. Note that there is
an abrupt change with $\phi$ in $\Delta s_1/\Delta \phi$ at the same
position of the peak of $s_2$, suggesting that the transition is of
second order.  We next show that the second order phase transition has
the same exponents as MF percolation. Figure~\ref{fig:ICOns} plots the
finite cluster size distribution of $\sigma_+$ agents, $n_s$ as a
function of $s$ at $f=f_c(\phi)$, from where we obtain $\tau=5/2$.
From $s_2$ we also compute the exponent $\gamma$ and obtain $\gamma
\approx 1$ (not shown). These two exponents indicate that the ICO
model on random graphs belongs to the same universality class as MF
percolation.

\section{The NCO on Coupled Networks Model}\label{NCOCOUPLE}

\begin{figure}[ht]
\includegraphics[width=5cm,height=6cm]{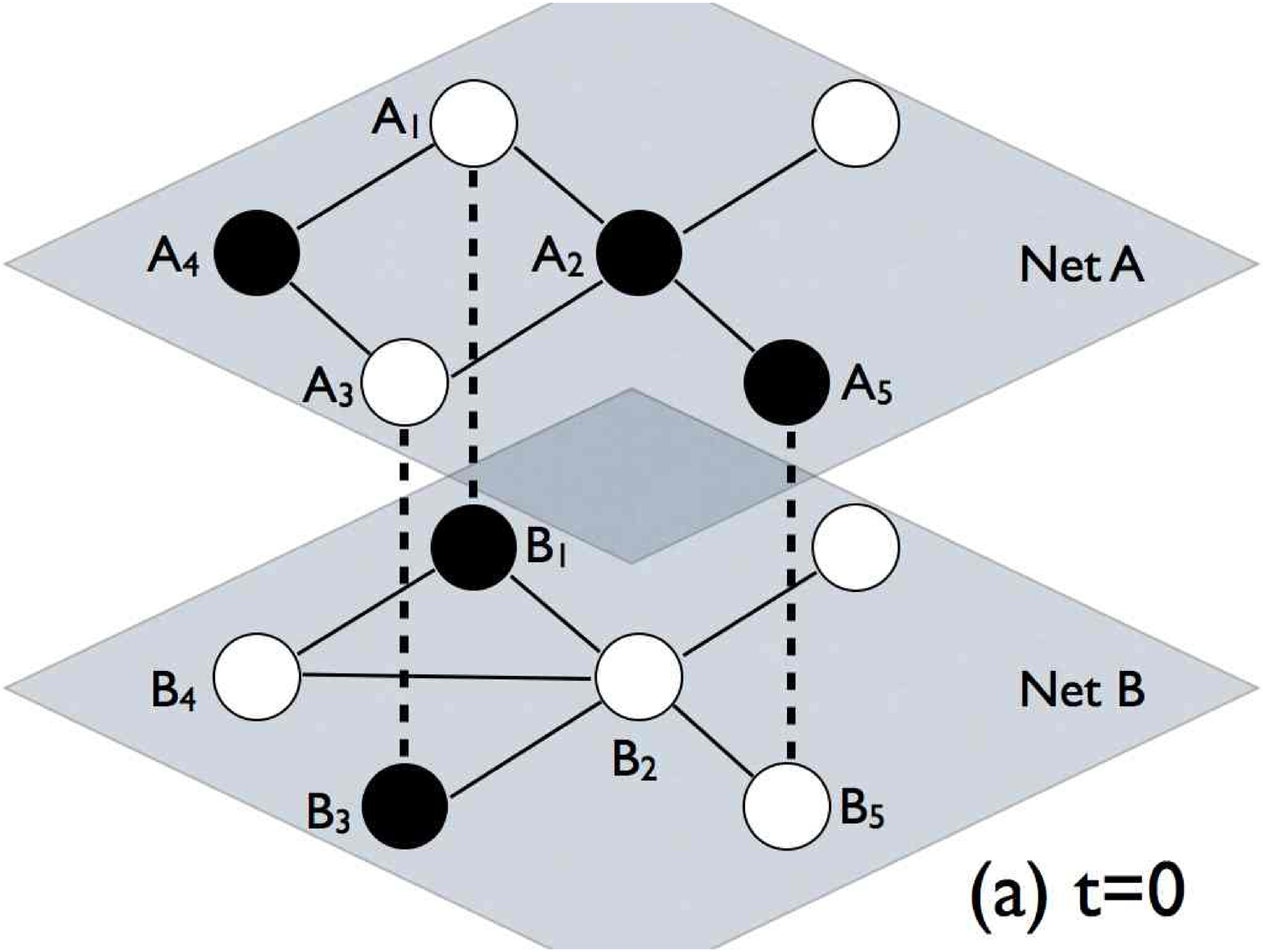}
\hspace{0cm}
\includegraphics[width=5cm,height=6cm]{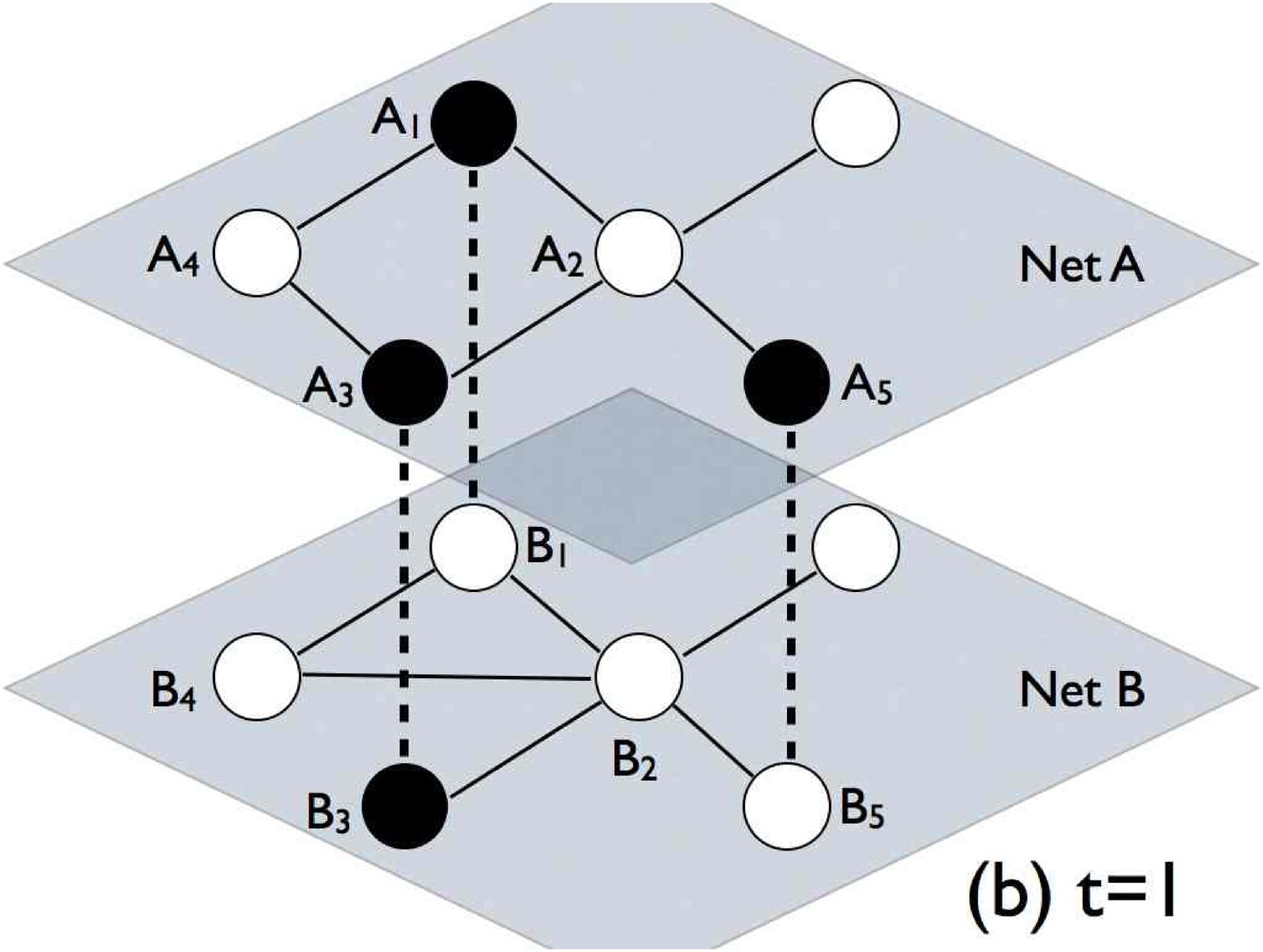}\\
\vspace{0cm}
\includegraphics[width=5cm,height=6cm]{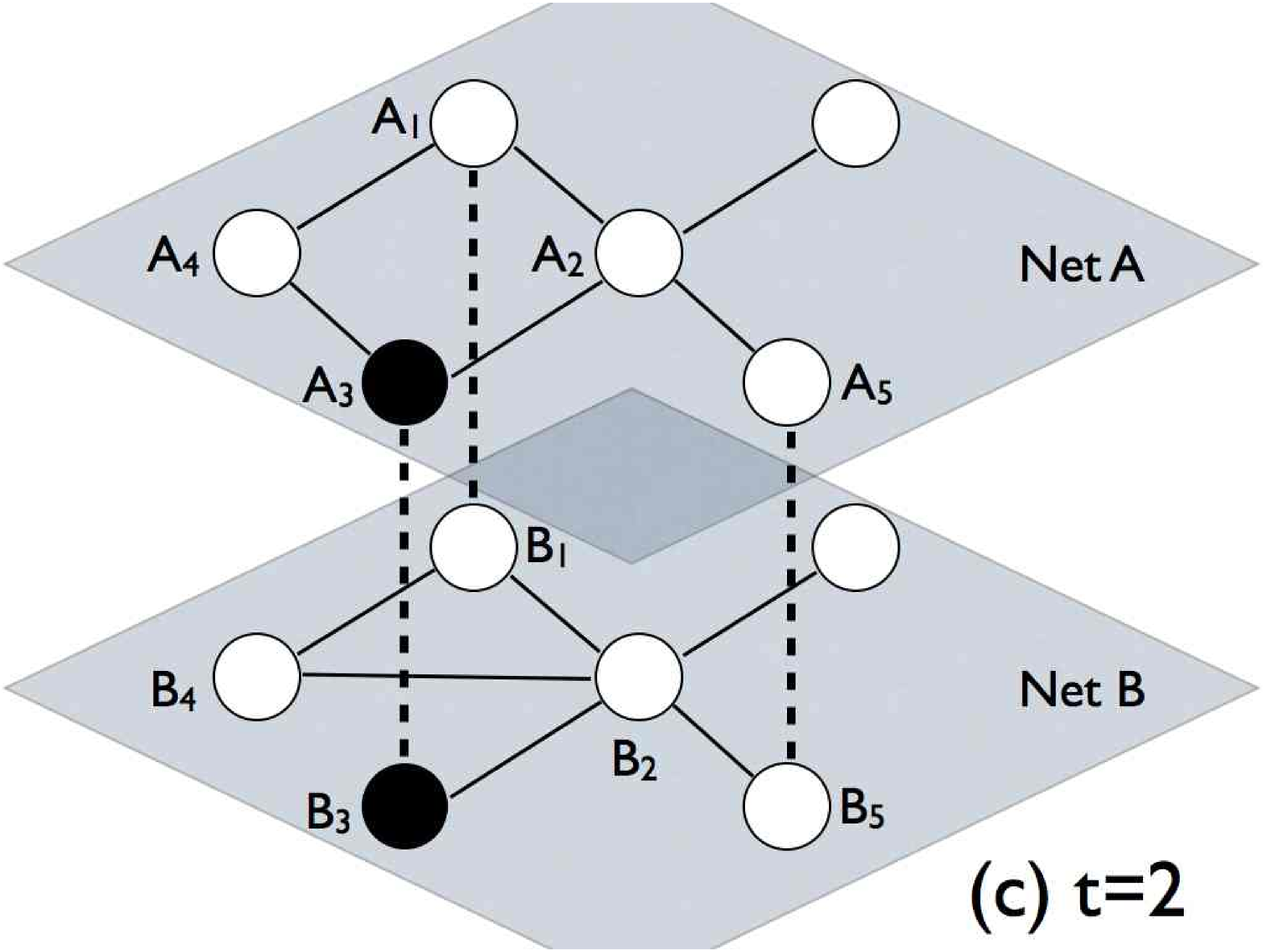}
\hspace{0cm}
\includegraphics[width=5cm,height=6cm]{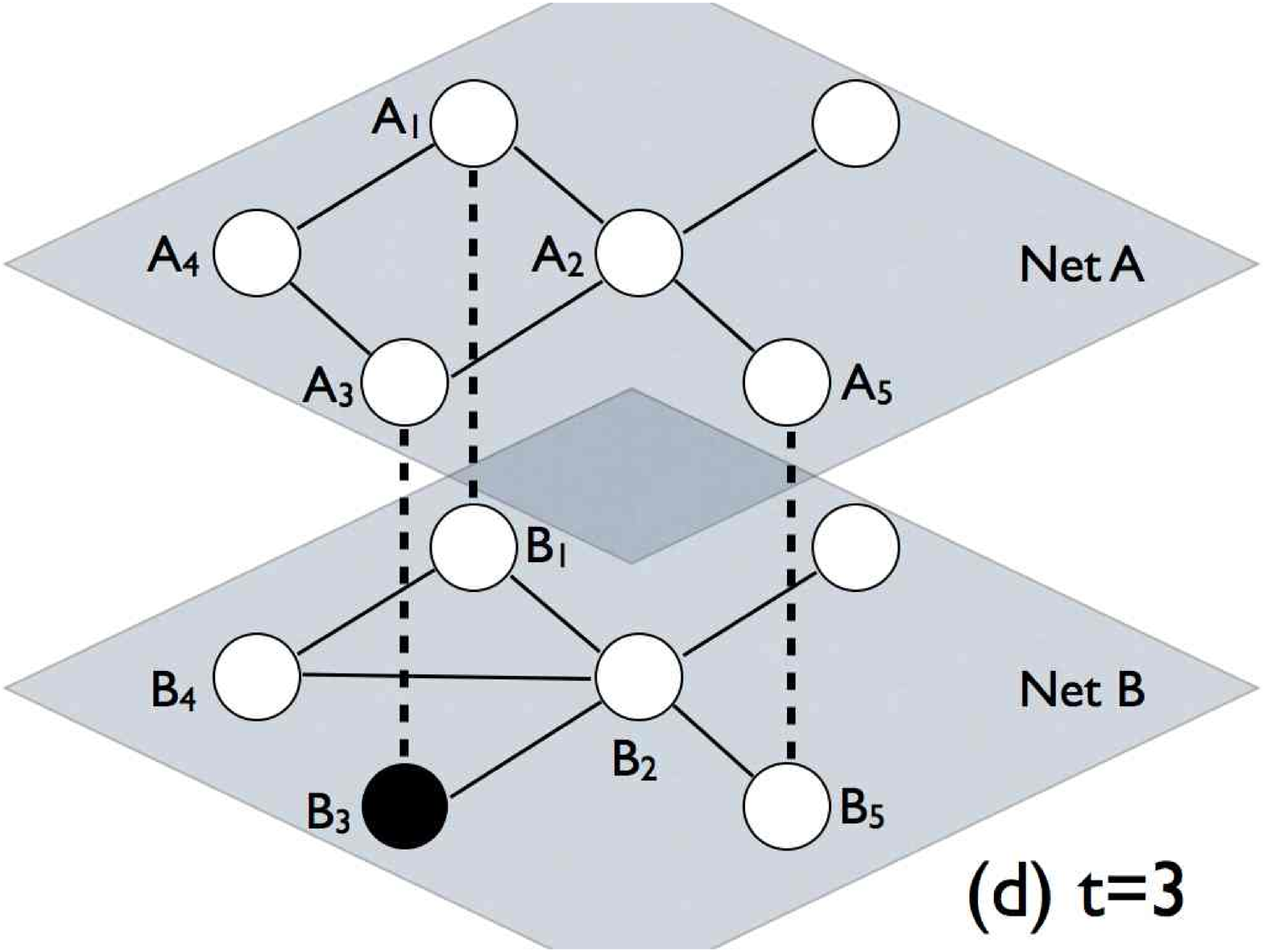}
\hspace{0cm}
\includegraphics[width=5cm,height=6cm]{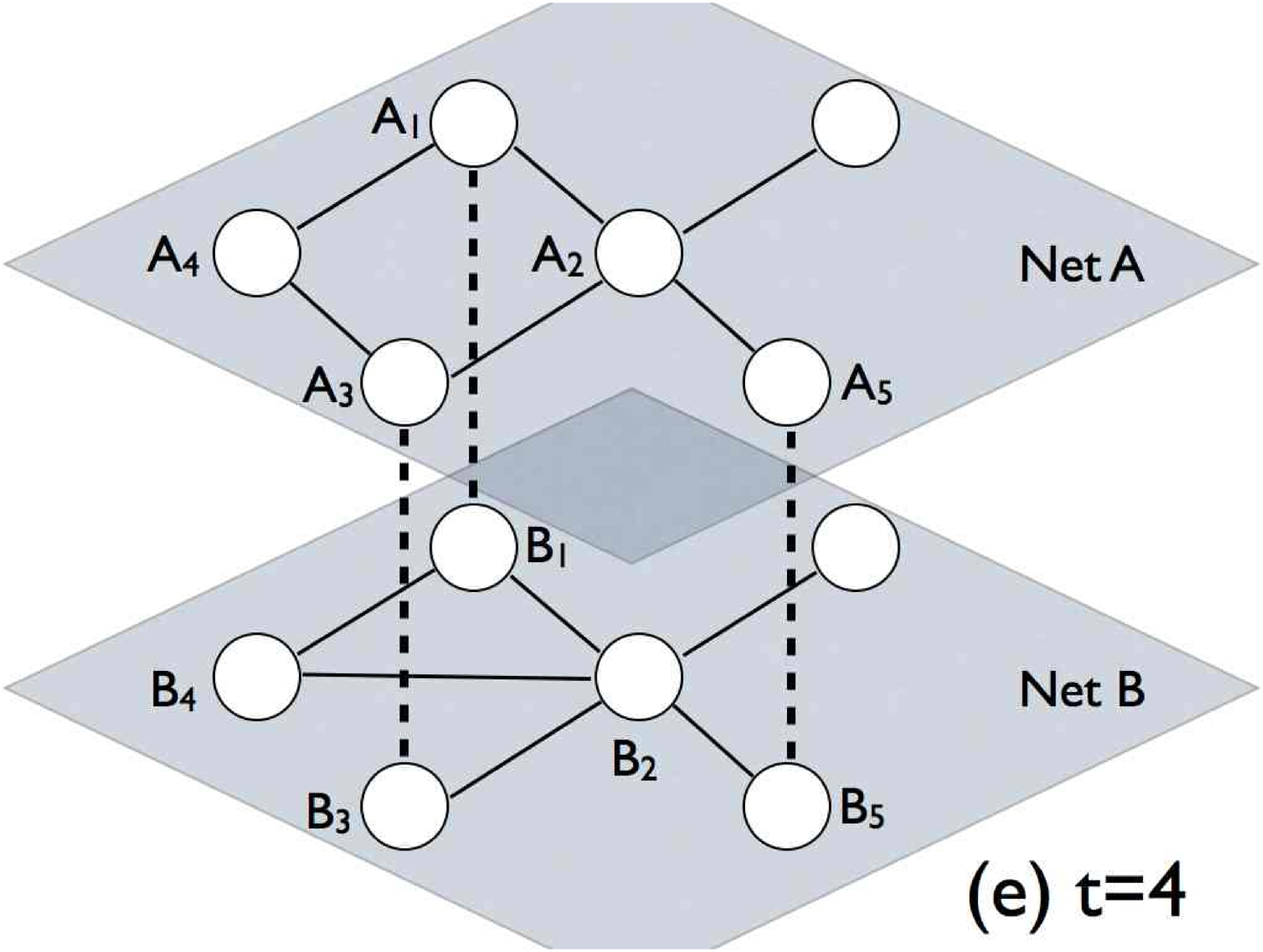}
\vspace{0cm}
\caption{(Color online) Schematic plot of the dynamics of the NCO on
  coupled networks showing the approach to a stable state on a system
  of interdependent networks $A$ and $B$ with $N=6$ nodes in each
  network. Open circles represent opinion $\sigma_+$ and solid circles
  represent opinion $\sigma_-$. The solid lines connecting nodes in
  each network are connectivity links within the networks, and the
  dashed lines connecting nodes from two networks are interdependent
  links. In the initial state, each node is randomly assigned with
  opinion $\sigma_+$ or $\sigma_-$. (a) At t=0, opinion dynamics
  evolve within each single network. In networks $A$ and $B$, nodes
  $A_1$, $A_2$, $A_3$, $A_4$ and $B_1$ are in a local minority opinion
  within their network, so at the end of this time step these nodes
  will change their opinions. The remaining nodes will keep their
  opinions. (b) At t=1, the two networks interact through the
  interdependent links. Notice that the global majority opinion is
  $\sigma_+$ now. Thus, the pairs $A_1 -B_1$ and $A_5 -B_5$ where two
  nodes have different opinions, will follow the global majority
  opinion. So at the end of this time step nodes $A_1$ and $A_5$ will
  change their opinions. The pair $A_3 -B_3$, remains as $\sigma_-$
  since both nodes share the same opinion. (c) At t=2, in network $A$,
  node $A_3$ is in a local minority, so at the end of the time step it
  will change opinion. (d) At t=3, the two networks interact through
  the interdependent links. Notice that the global majority opinion is
  still $\sigma_+$, the only pair with different opinion is $A_3
  -B_3$, so at the end of this time step, $B_3$ will change its
  opinion. (d) At t=4, the interdependent networks reach a stable
  state and no more changes will happen.
\label{fig:f0}}
\end{figure}

Figure~\ref{fig:f0} demonstrates the dynamics of the NCO model on
coupled networks. In coupled networks, the two networks represent two
groups of people. The links within each network denote the
relationships between nodes. For simplicity, we assume that the two
networks have the same number of nodes $N$ and the same degree
distribution. We also assign these two networks the same initial
opinion condition, i.e., in both networks there is initially a
fraction $f$ of nodes holding the $\sigma_+$ opinion, and a fraction
$1-f$ holding the $\sigma_-$ opinion. To represent the strong social
coupling between the two groups, we randomly choose a fraction $q$ of
the nodes from both networks to form $qN$ pairs of one-to-one
interdependent pairs regardless of their original opinions. At time
t=0, in both networks we apply the same opinion formation rule, the
NCO model, to decide whether an agent will change its opinion
regardless of the interdependent links. This means that at this stage
opinions propagate in each single network independently---as though
the other network does not exist. All opinion updates are made
simultaneously and in parallel. At t=1, if two nodes with an
interdependent link have the same opinion they keep that opinion. If
they do not, they follow the majority opinion of the coupled networks
(global majority rule). All interdependent agents update their
opinions simultaneously at the end of this time step. We repeat these
two steps until the system reaches a steady state.

\subsection{Simulation Results}

We perform simulations of the NCO on coupled networks where both of
the interdependent networks are either ER networks with $\langle k
\rangle=4$ or SF networks with $k_{\rm min}=2$ and $\lambda=2.5$. For
an initial fraction $f$ of opinion $\sigma_+$ and a fraction $q$ of
interdependent links, the NCO on coupled networks is simulated on
$10^4$ network realizations to explore how interdependent links affect
opinion dynamics.

\subsubsection{NCO on Coupled ER Networks}

\begin{figure}[ht]
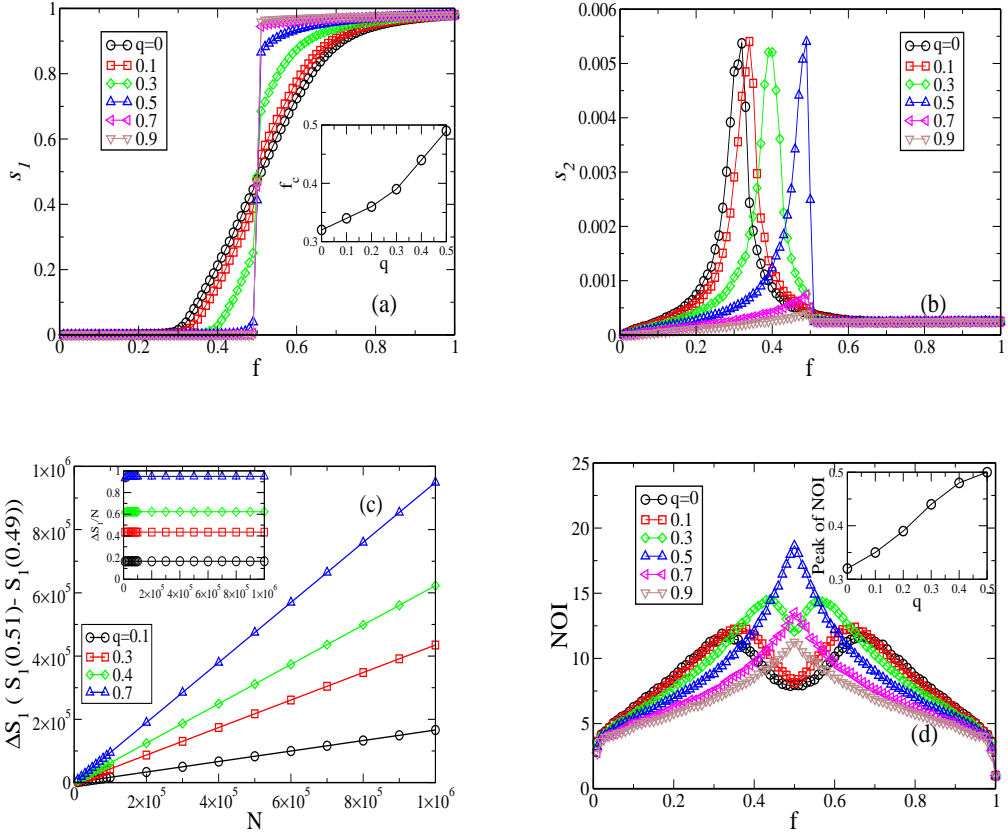

\includegraphics[width=6cm,height=5cm]{fig1_a1.eps}
\hspace{1cm}
\includegraphics[width=6cm,height=5cm]{fig1_b1.eps}\\
\vspace{1cm}
\includegraphics[width=6cm,height=5cm]{DeltaS1_N_1.eps}
\hspace{1cm}
\includegraphics[width=6cm,height=5cm]{fig1_c1.eps}
\vspace{0cm}
\newpage
\caption{Plots of NCO on coupled ER networks, with $\langle
  k_A\rangle=\langle k_B\rangle=4$ and for each network $N=10000$. (a)
  Plot of $s_1$ of opinion $\sigma_+$ as a function of $f$ for several
  values of $q$. As seen, when $q=0$, the system undergoes a smooth
  second order phase transition (regular NCO model). As $q$ increases
  until $q=0.5$, it becomes a hybrid phase transition, which contains
  both a smooth second order type and a seemingly abrupt jump, {\it
    i.e.}, $s_1$ changes smoothly close to $f_c(q)$ and followed by a
  sharp jump at $f=0.5$. When $q$ is further increased ($q>0.5$), the
  smooth phase transition disappears, the system undergoes a pure
  abrupt phase transition. In the inset of (a) we plot $f_c$ as a
  function of $q$. (b) Plot of $s_2$ as a function of $f$ for
  different values of $q$. As seen, when $q$ increases, the peaks of
  the $s_2$, which characterize the critical threshold value of the
  second order phase transition, shift to the right. We can see that
  beyond $q=0.5$, the peak of $s_2$ disappears, which indicates that
  there is no second order phase transition. (c) Plot of the change of
  $S_1$ around $f=0.5$, $\Delta S_1$, as a function of system size $N$
  for different values of $q>0$. In the inset of (c), we plot $\Delta
  S_1/N$ as a function of $N$ for different values of $q$. The linear
  relationship between $\Delta S_1$ and $N$ suggests that for $q>0$,
  around $f=0.5$, there exists a discontinuous transition. (d) Plot of
  the number of cascading steps (NOI) of the networks as a function of
  $f$ for different values of $q$. As seen there are two peaks for
  each $q$ value for $q<0.5$, and for $q \ge 0.5$ there is only one peak
  at $f=0.5$. In the inset of (d), plot of the location of NOI peak as
  a function of $q$. The solid lines are guides for the
  eyes. \label{fig:f1}}
\end{figure}

We first investigate $s_1$ as a function of $f$ for different values
of $q$. Figure~\ref{fig:f1}(a) shows that when $q=0$, which
corresponds to the NCO model on a single network, the system undergoes
a second order phase transition with a critical threshold $f_c$
\cite{NCO}. When $q>0$, there are two regions $0 < q \le 0.5$ and $q >
0.5$. For the region $0 < q \le 0.5$, as in the NCO model on a single
network, the second order phase transition still exists, but the
critical value $f \equiv f_c(q)$ is increasing with $q$. The value of
$f_c(q)$ can be determined by the location of the peak of $s_2$, which
is shown in Fig.~\ref{fig:f1}(b), where we plot $s_2$ as a function of
$f$ for different values of $q$.  The inset of Fig.~\ref{fig:f1}(a)
shows a plot of $f_c(q)$ as a function of $q$. We find that the peak
of $s_2$ shifts to the right for $q \le 0.5$ as $q$ increases, which
means that $f_c(q)$ increases as $q$ increases. This suggests that if
we add more interdependent links between the two networks, the
minority opinion will need a larger initial fraction in order to
exist. In the region $0 < q \le 0.5$ we also find that, unlike the NCO
model on a single ER network, there is an abrupt change of $s_1$ at
$f=0.5$, indicating that in addition to the smooth second order phase
transition at $f_c(q)$, there may also be a discontinuous transition
at $f=0.5$. Our results suggest that when $0 < q \le 0.5$ the system
may undergo a hybrid phase transition \cite{hybrid}, which is a
mixture of both an abrupt and a second order phase transition. We also
find that as $q$ increases the discontinuity around $f=0.5$ becomes
more pronounced. Although the system possesses a seemingly
discontinuous phase transition for $0 < q \le 0.5$ , the model itself
is still a non-consensus model, i.e., when $f$ is above the critical
value $f_c(q)$ the two opinions coexist in a steady state. When $q >
0.5$ the smooth second order phase transition of $s_1$ disappears and
is replaced by an abrupt transition at $f=0.5$. When $q > 0.5$ the
peak of $s_2$ disappears, supporting the loss of the second order
phase transition [see Fig.~\ref{fig:f1}(b)], and the system undergoes
a pure abrupt transition. This suggests that when interactions between
networks are sufficiently strong the hybrid phase transition is
replaced by a pure abrupt transition. For all values of $q$, the
region where two opinions can coexist decreases as $q$ increases, and
the NCO coupled networks model moves at large $q$ toward the consensus
type opinion model. To further support our finding of the existence of
a discontinuous transition when $q>0$, in Fig.~\ref{fig:f1}(c) and its
inset we plot respectively the jump of $S_1$, $\Delta
S_1=S_1(0.51)-S_1(0.49)$, and $\Delta S_1/N$, around $f=0.5$ as a
function of the system size $N$ for different values of $q$.  The
linear relationship between $\Delta S_1$ and $N$ supports our
assumption of the existence of a discontinuous transition around
$f=0.5$ for all values of $q>0$. Note that, as the value of $q$
increases, the value of $\Delta S_1$ increases, which means that as we
increase the value of $q$ the abrupt transition becomes more
pronounced.

To further support our conclusions, we investigate the number of
iterations (NOI), which is the number of time steps needed to reach
the steady state, as a function of $f$ for different values of
$q$. Figure~\ref{fig:f1}(d) shows a plot of the NOI as a function of
$f$ for different values of $q$. As described in
Ref.~\cite{ParshaniPNAS}, in a pure first order phase transition due
to cascading failures the location of the peak of the NOI determines
the critical threshold of the transition, which is the case for
$q>0.5$ in our model.  Figure~\ref{fig:f1}(d) shows that there is only
one peak for the NOI curves for $q>0.5$ at $f=0.5$, which is the
position of the critical threshold of the abrupt transition. In the
hybrid phase transition for $q \le 0.5$, the relation between the peak
of the NOI and the critical threshold is unclear because there are two
critical thresholds, one for the discontinuous transition at $f=0.5$
and the other for the second order phase transition at
$f_c(q)$. Figure~\ref{fig:f1}(d) shows that when $q<0.5$ the NOI
exhibits two symmetric peaks. The inset of Fig.~\ref{fig:f1}(d) shows
a plot of the location of the left peaks of the NOI as a function of
$q$. Comparing the insets in Fig.~\ref{fig:f1}(a) and
Fig.~\ref{fig:f1}(d), we find that the curve of the peak locations of
the NOI is always above the $f_c(q)$ curve, which suggests that for a
hybrid phase transition the peak of the NOI is located between the
critical thresholds of the second order phase transition and the
abrupt transition.

\begin{figure}[ht]
\includegraphics[width=14cm,height=14cm]{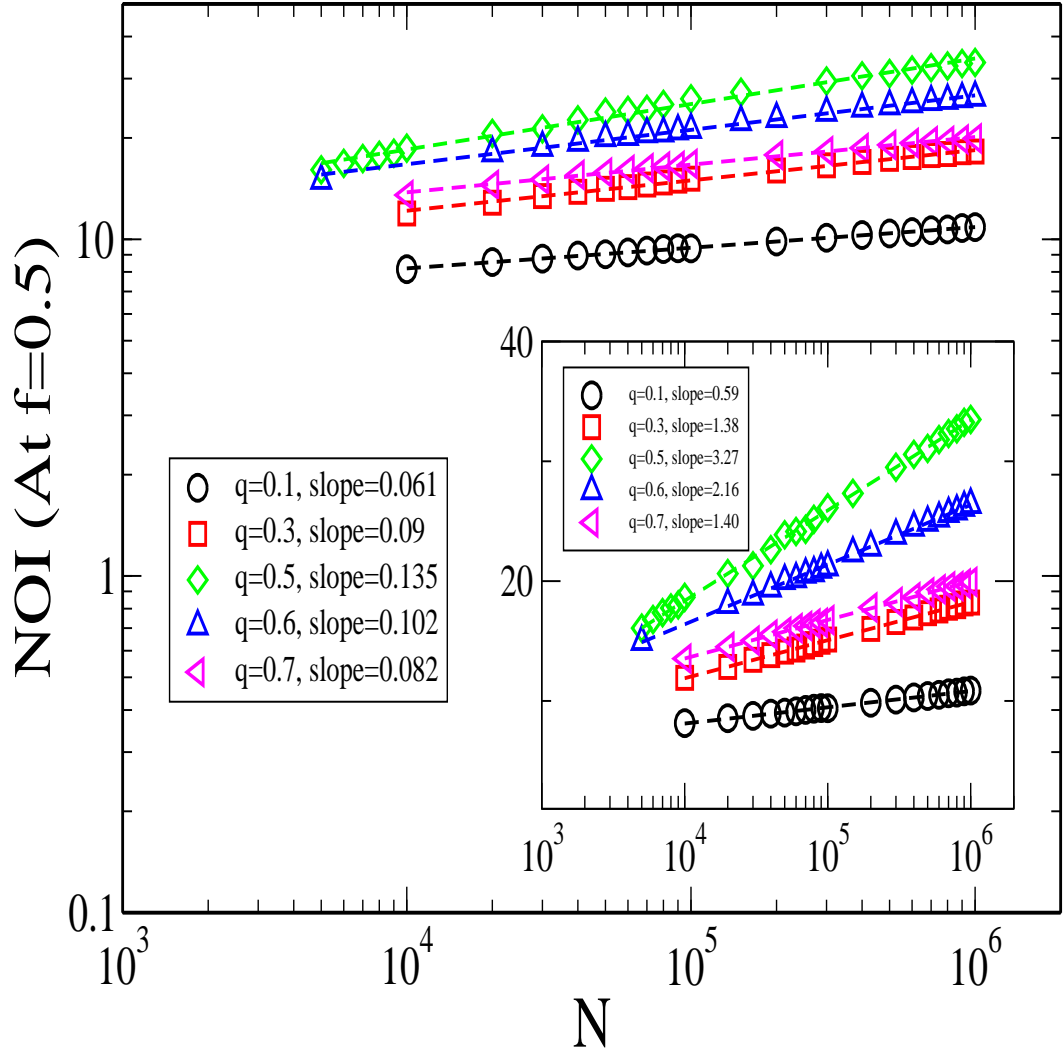}\\
\vspace{1cm}
\caption{ Plot of the number of cascading steps between the networks (ER
  with $\langle k_A\rangle=\langle k_B\rangle=\langle k\rangle=4$) at
  $f=0.5$ as a function of system size for different values of $q$ in
  log-log scale and in log-linear scale in the inset respectively. The
  dashed lines are the power law and logarithmic fittings
  respectively. \label{fig:f2}}
\end{figure}

Figure~\ref{fig:f2} shows a log-log plot of the NOI at $f=0.5$ as a
function of the system size $N$ for different values of $q$, and the
inset of Fig.~\ref{fig:f2} shows the same in a log-linear plot. The
accuracy of the simulations is such that we cannot distinguish the
relationship between exponential and logarithmic. However, the
increase of NOI with system size indicates that there is a real jump
at approximately $f=0.5$ rather than a finite size effect. This
supports our previous conjecture that for all values of $q>0$, the NCO
on coupled networks exhibits an abrupt transition at $f=0.5$.

\begin{figure}[ht]
\includegraphics[width=14cm,height=14cm]{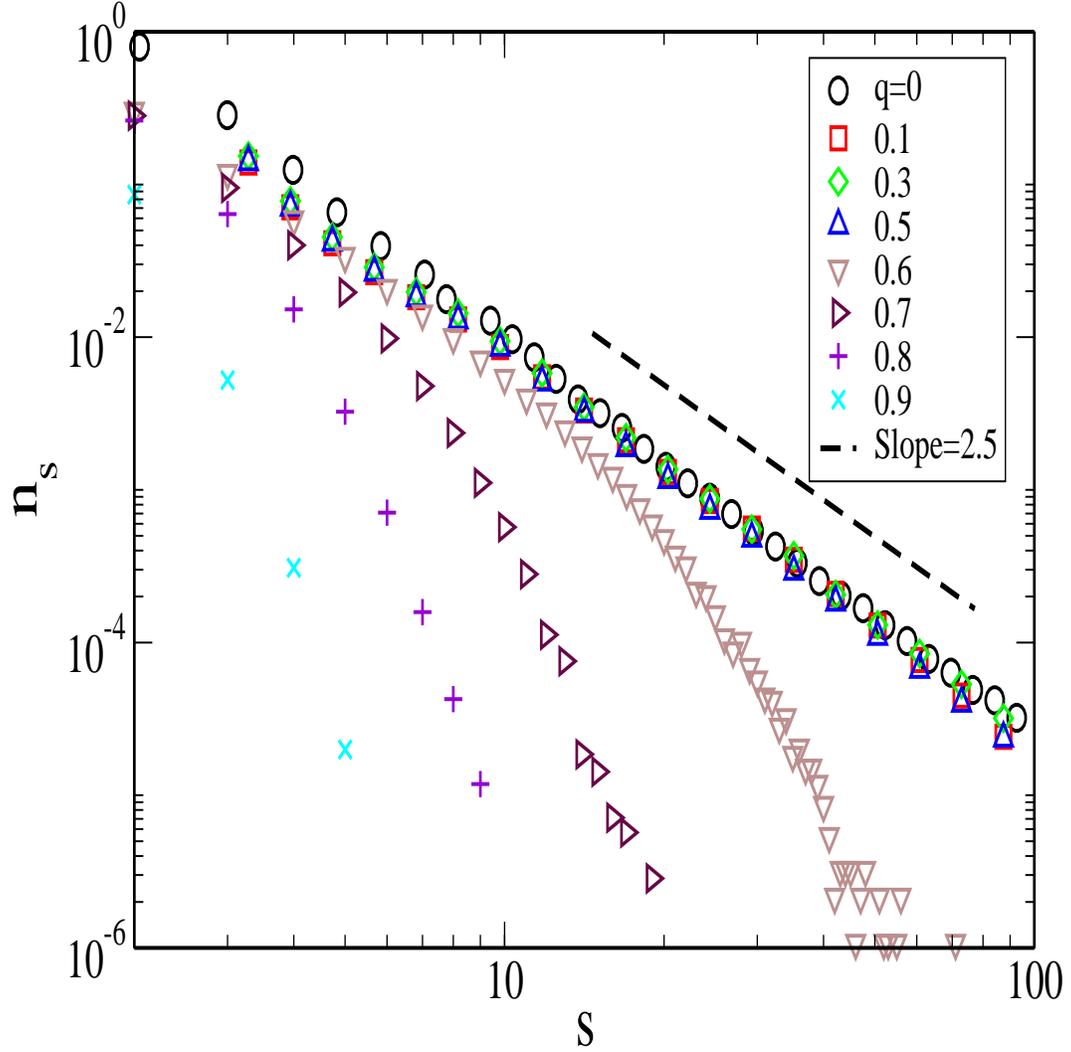}\\
\vspace{1cm}
\caption{ Plot of $n_s$ as a function of $s$ at criticality, $f_c$,
  for different values of $q$ for the NCO model on coupled ER
  networks, with $\langle k_A\rangle=\langle k_B\rangle=\langle
  k\rangle=4$ and $N=10000$. For each network the results are averaged
  over $10^4$ realizations. As $q$ increases, $n_s$ losses the power
  law shape indicating that the second order phase transition is
  lost. The dashed line is a guide to show the slope
  $\tau=2.5$.\label{fig:f3}}
\end{figure}

We next present results indicating that, when $q \le 0.5$ and when $f$
is close to $f_c(q)$, our model is in the same universality class as
regular MF percolation, even though a discontinuity appears at larger
$f$. For regular percolation on random graphs at criticality, the
cluster sizes follow a power law distribution, $n_s \sim s^{-\tau}$
with $\tau=2.5$ \cite{percolation,quenched_shlomo, Complexnetwork}.
Figure~\ref{fig:f3} shows a plot of $n_s$ as a function of $s$ for
finite $\sigma_+$ clusters at criticality, $f_c(q)$. We see that for
$q \le 0.5$, $\tau \approx 2.5$, and for $q>0.5$, the power law no
longer holds. The exponent values we obtain strongly indicate that,
for small value of $q$, the NCO model on coupled ER networks close to
$f_c$ is in the same universality class as mean field percolation in
random networks.  The power law for the cluster size distribution at
$q>0.5$ disappears, so we conclude that the NCO coupled networks model
changes the phase transition type as $q$ increases from $q \le 0.5$ to
$q>0.5$.

\subsubsection{NCO on Coupled SF Networks} 

Empirical studies show that many real-world social networks are not
ER. They instead exhibit a SF degree distribution \cite{SF1} in which
$P(k)\sim k^{-\lambda}$ and $\lambda$ characterize the broadness of
the distribution. A feature of SF is the existence of hubs, i.e., very
high degree nodes. These large hubs make the opinion dynamic
processes in SF networks much more efficient than in ER networks
\cite{SFbetter1,SFbetter2,SFbetter3,SFbetter4,SFbetter5,SFbetter6}.

\begin{figure}[ht]
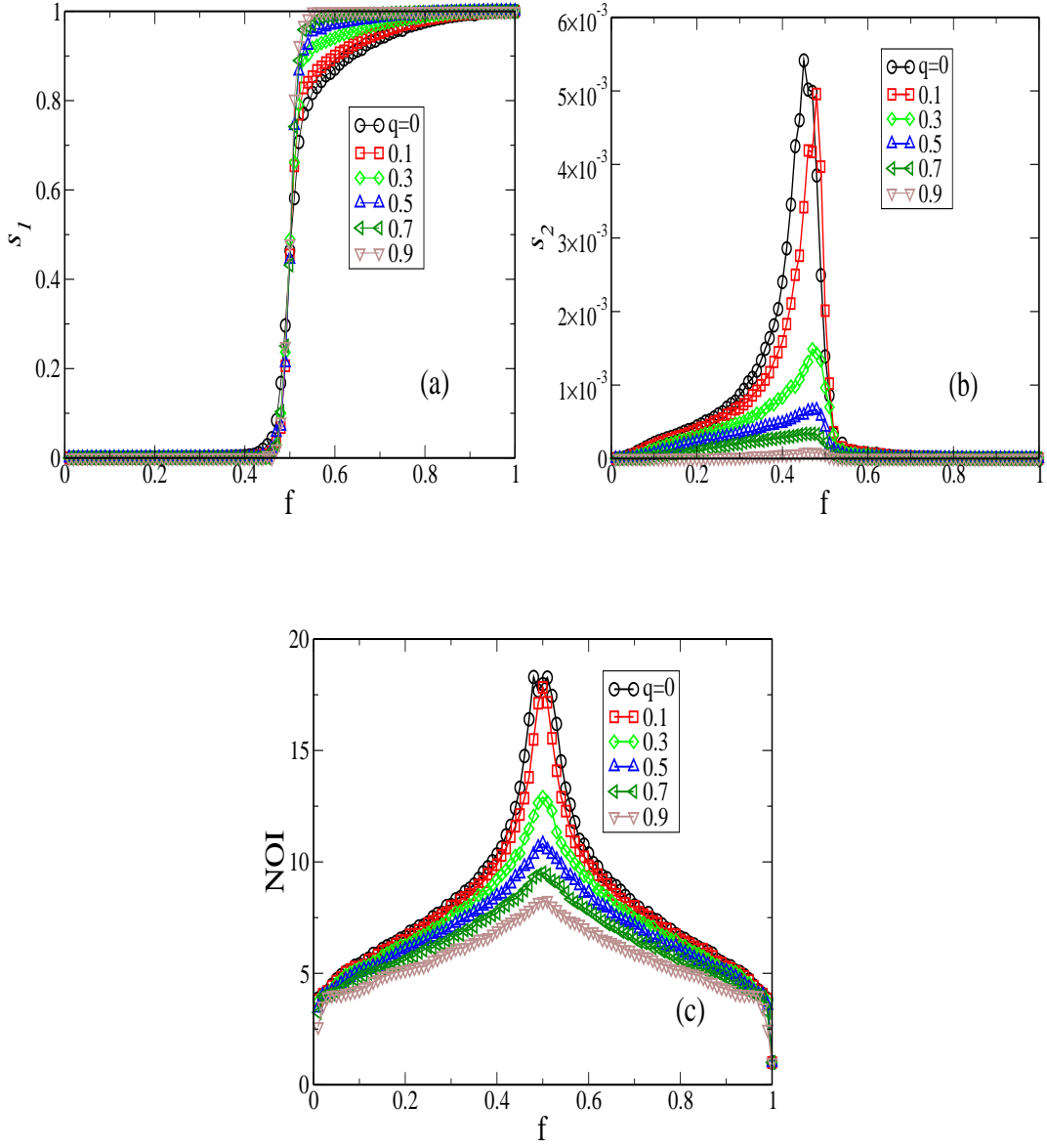

\includegraphics[width=7cm,height=7cm]{fig4_a.eps}
\includegraphics[width=7cm,height=7cm]{fig4_b.eps}\\
\vspace{1.5cm}
\includegraphics[width=7cm,height=7cm]{fig4_c.eps}
\vspace{1cm}
\caption{Study of NCO model on coupled SF networks, with $k_{\rm
    min}=2$, $\lambda=2.5$ and $N=10000$ for each network. (a) Plot of
  $s_1$ of opinion $\sigma_+$ as a function $f$ for different values
  of $q$. (b) Plot of $s_2$ as a function of $f$ for different values
  of $q$. (c) Plot of the number of cascading steps, NOI, of the
  coupled networks as a function of $f$ for different values of
  $q$. The solid lines are guides for the eyes.\label{fig:f4}}

\end{figure}

\begin{figure}[ht]
\includegraphics[width=14cm,height=14cm]{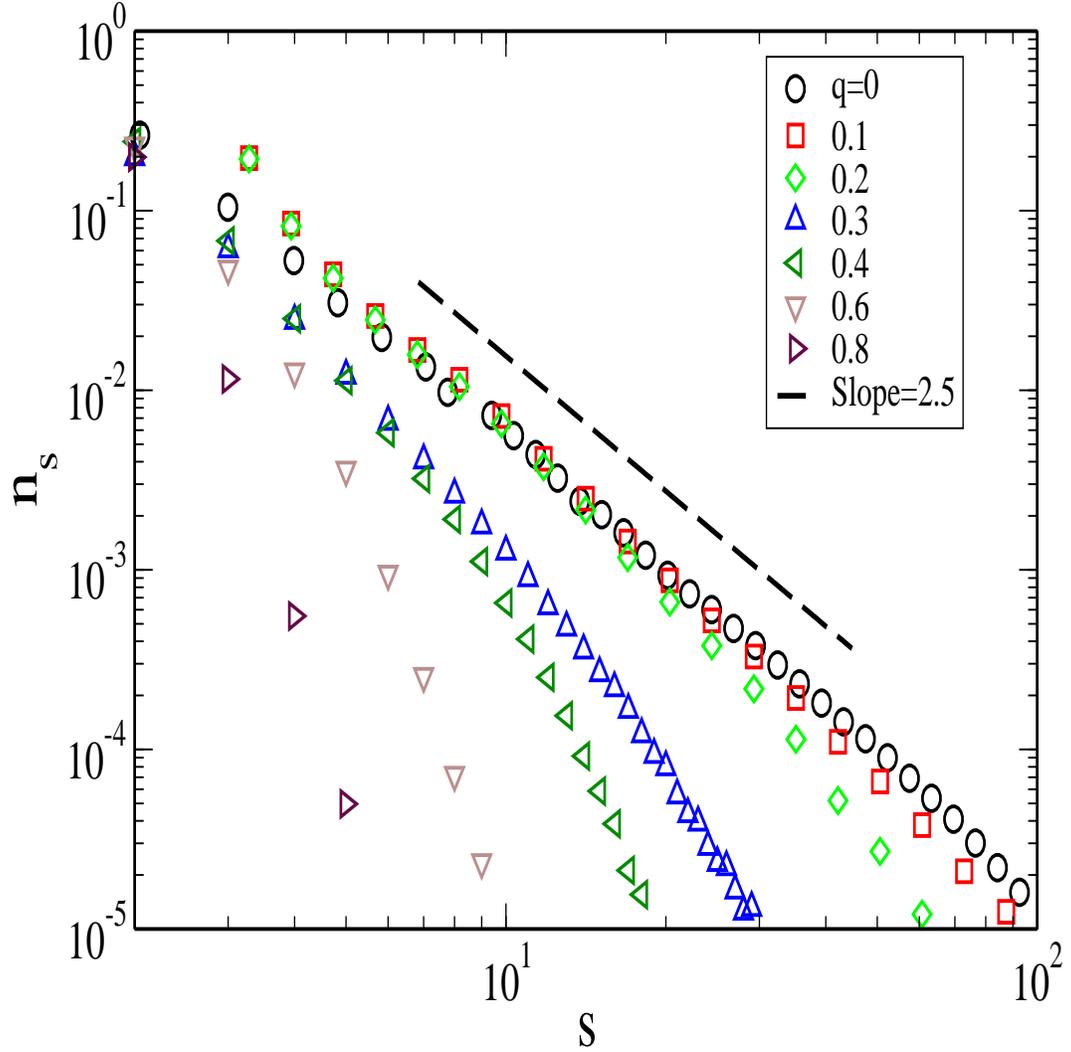}\\
\vspace{1cm}
\caption{ Plot of $n_s$ as a function of $s$ at criticality for
  different value of $q$ for the NCO model on coupled SF networks,
  with $k_{\rm min}=2$, $\lambda=2.5$ and $N=10000$. For each case the
  results are averaged over $10^4$ realizations. As $q$ increases,
  $n_s$ losses the power law shape indicating that the second order
  phase transition is lost. The dashed line is a guide to show the
  slope $\tau=2.5$.\label{fig:f5}}
\end{figure}

Because of its large number of connections, a hub in the NCO model
tends to follow the opinion of the majority and effectively influence
the opinions of its neighbors. In a SF network the hubs help the
majority dominate the minority, and thus the NCO model on a single SF
network has a larger $f_c$ and exhibits a much sharper jump around
$f_c$ than in ER networks with the same average degree
\cite{NCO}. This is also the case in interdependent SF
networks. Figures~\ref{fig:f4}(a), \ref{fig:f4}(b), and
\ref{fig:f4}(c) depict $s_1$, $s_2$, and NOI as a functions of $f$ for
different values of $q$, respectively. The results of the NCO model on
coupled SF networks are similar to those of coupled ER networks,
except that the region of the hybrid phase transition is much smaller
in coupled SF networks. This is confirmed by the fact that the peak of
$s_2$ drops much faster for small $q$ values and that the single peak
of NOI shows up at smaller $q$ values for coupled SF networks in
contrast to the case of coupled ER networks. This indicates that the
pure abrupt phase transition occurs at smaller $q$ values in coupled
SF networks compared to coupled ER networks, which suggests that in
coupled SF networks a smaller number of interdependent agents are
needed to achieve a consensus state compared to coupled ER networks.
Figure~\ref{fig:f5} shows a plot of $n_s$ as a function of $s$ for
finite $\sigma_+$ clusters at criticality. Note that in SF networks
when $q \leq 0.1$ the $n_s$ decays as a power law with $\tau=2.5$, and
when $q>0.1$ the power law decay of $n_s$ no longer holds. This
suggests that only for small values of $q$ our NCO model on coupled SF
networks is in the same universality class as regular MF
percolation. Comparing Fig.~\ref{fig:f5} with Fig.~\ref{fig:f3}, we
find that the power law decay disappears at smaller $q$ values in
coupled SF networks compared to coupled ER networks. This supports our
hypothesis that interdependent links push the entire system to an
abrupt phase transition more effectively in coupled SF networks than
in coupled ER networks.

In both coupled ER and SF networks, our non-consensus opinion second
order phase transition model is transformed into a consensus opinion
type abrupt transition model when the number of interdependent links
is increased. This suggests that increasing the interactions between
different groups in our world will push humanity to become
increasingly homogeneous, i.e., interdependent pairs in the NCO
coupled networks model helps the majority opinions supporters to
eliminate the minority opinion, making uniformity (consensus) a
possible final result.

\section{Summary}\label{conclusion}

In this paper we revisit and extended the non-consensus opinion (NCO)
model, introduced by Shao et al. \cite{NCO}. We introduce the NCO$W$
model in which each node's opinion is given a weight $W$ to represent
the nodes' resistance to opinion changes. We find that in both the NCO
and the NCO$W$ models the size of the largest minority cluster with
$\sigma_+$ opinion undergoes a second order MF percolation transition
in which the control parameter is $f$. The NCO$W$ model is more robust
than the NCO model because the weighted nodes reinforce the largest
$\sigma_+$ minority cluster, and shift the critical value $f_c$ to
values lower than those found in the NCO model.  We also show that
when the average network degree $\langle k\rangle$ in the NCO model is
increased, the second order phase transition is replaced by an abrupt
transition, making the NCO model converge to a consensus type opinion
model. We also review another non-consensus opinion model, the ICO
model \cite{ICO}, which introduces into the system, using both random
and targeted strategies, a fraction $\phi$ of inflexible contrarians
(which act as quenched noise). As $\phi$ increases, both random and
targeted strategies reduce the size of the largest $\sigma_+$ cluster
and, above a critical threshold $\phi=\phi_c$, the largest $\sigma_+$
cluster disappears and the second order phase transition is also
lost. The targeted strategy is more efficient in eliminating the
largest $\sigma_+$ cluster or decreasing its size. This is due to the
fact that the contarians are introduced (targeted) mainly into the
largest cluster, which contains most of the high degree nodes. Thus a
smaller $\phi_c$ value is needed to eliminate the largest cluster of
minority in the targeted strategy compared to the random strategy. We
also study an opinion model in which two interdependent networks are
coupled by a fraction $q$ of interdependent links. The internal
dynamics within each network obey the NCO rules, but the cross-network
interdependent nodes, when their opinions differ, obey the global
majority rule. These interdependent links force the system from a
second order phase transition, characteristic of the NCO model on a
single network, to a hybrid phase transition, i.e., a mix of a second
order transition and an abrupt transition. As the fraction of
interdependent links increases, the system evolves to a pure abrupt
phase transition. Above a certain value of $q$, which is strongly
dependent on network topology, the interdependent link interactions
push the non-consensus opinion model to a consensus opinion model.
Because scale free networks have large hubs, the effect of
interdependent links is more pronounced in interdependent scale free
networks than in interdependent Erd\"{o}s R\'{e}nyi networks. We are
investigating whether the same effect appears in other opinion models
of interdependent networks. The results will be presented in a future
paper.

We acknowledge support from the DTRA, ONR, NSF CDI program, the
European EPIWORK, LINC and MULTIPLEX projects, the Israel Science
Foundation, the PICT 0293/00 and UNMdP, NGI and CONGAS.

%\bibliographystyle{unsrt}
%\bibliography{bibpaper_corr.bib}

\newpage
\newpage

\newpage

\newpage

\newpage


\end{document}